\documentclass[10pt,journal,compsoc]{IEEEtran}
\usepackage{array,multirow,makecell}
\usepackage[T1]{fontenc}
\usepackage{graphicx}
\usepackage{xcolor}
\usepackage{booktabs}
\usepackage{cite}
\usepackage{amsmath}
\usepackage{url}
\usepackage{algpseudocode}
\usepackage{enumitem}
\usepackage{listings}
\usepackage[utf8]{inputenc}
\usepackage{tabularray}
\usepackage{empheq}
\usepackage[most]{tcolorbox}
\tcbuselibrary{listings}
\tcbuselibrary{skins}
\usepackage{longtable}

\usepackage{algorithm}
\usepackage{colortbl}
\usepackage{pgfplots}
\pgfplotsset{compat=1.18}
\usepackage{amssymb}
\usepackage{pifont}

\usepackage{tikz}
\usepackage{float} 
\usetikzlibrary{patterns}
\usetikzlibrary{arrows,shapes,positioning,shadows,trees}
\usepackage{rotating, adjustbox}
\usepackage{hyperref}
\usepackage{balance}
\usepackage{placeins}
\usepackage{orcidlink}
\DeclareMathOperator*{\argmax}{arg\,max}

\newcommand{\method}{VulnScout-C}

\title{\method: A Lightweight Transformer for C Code Vulnerability Detection}

\author{Aymen Lassoued\,\orcidlink{0009-0002-9807-2901}\IEEEauthorrefmark{1},
Nacef Mbarek\,\orcidlink{0009-0000-0688-3164}\IEEEauthorrefmark{1},
Bechir Dardouri\,\orcidlink{0009-0003-3287-5469}\IEEEauthorrefmark{1},
Bassem Ouni\,\orcidlink{0000-0001-6534-9295}\IEEEauthorrefmark{2}\thanks{Corresponding author: Bassem Ouni (bassem.ouni@ku.ac.ae).},
Qing Li\,\orcidlink{0000-0002-6442-5003}\IEEEauthorrefmark{3},
and Fakhri Karray\,\orcidlink{0000-0002-4217-1372}\IEEEauthorrefmark{4}%
\IEEEcompsocitemizethanks{%
\IEEEcompsocthanksitem \IEEEauthorrefmark{1}Ecole Polytechnique de Tunisie, La Marsa, Tunisia. E-mail: \{aymen.lassoued, nacef.mbarek, bechir.dardouri\}@ept.ucar.tn.%
\IEEEcompsocthanksitem \IEEEauthorrefmark{2}Khalifa University, Abu Dhabi, UAE. E-mail: bassem.ouni@ku.ac.ae.%
\IEEEcompsocthanksitem \IEEEauthorrefmark{3}University of Groningen, The Netherlands. E-mail: qing.li@rug.nl.%
\IEEEcompsocthanksitem \IEEEauthorrefmark{4}Mohamed bin Zayed University of Artificial Intelligence (MBZUAI), Abu Dhabi, UAE. E-mail: fakhri.karray@mbzuai.ac.ae.%
}%
}

\begin{document}
\markboth{IEEE Transactions on Dependable and Secure Computing,~Vol.~XX, No.~X, Month~20XX}%
{Lassoued \MakeLowercase{\textit{et al.}}: \method: A Lightweight Transformer for C Code Vulnerability Detection}
\maketitle
\begin{abstract}
Vulnerability detection in C programs is a critical challenge in
software security. Although large language models (LLMs) achieve
strong detection performance, their multi-billion-parameter scale
makes them impractical for integration into development workflows
requiring low latency and continuous analysis.

We introduce \textsc{VulnScout-C}, a compact transformer
architecture with 693M total parameters (353M active during
inference), derived from the Qwen model family and optimized for
C code vulnerability detection. Alongside the model, we present
\textsc{VulnScout}, a new 33,565-sample curated dataset generated
through a controlled multi-agent pipeline with formal verification,
designed to fill coverage gaps in existing benchmarks across
underrepresented CWE categories. Evaluated on a standardized
C vulnerability detection benchmark, \textsc{VulnScout-C}
outperforms all evaluated baselines, including state-of-the-art
reasoning LLMs and commercial static analysis tools, while
offering a fraction of their inference cost. These results
demonstrate that task-specialized compact architectures can
match or even outperform the detection capability of models orders of magnitude
larger, making continuous, low-latency vulnerability analysis
practical within real-world development workflows.
\end{abstract}


\section{Introduction}
\label{sec:intro}

\IEEEPARstart{S}{oftware} vulnerabilities represent a persistent threat to system security, with memory-related errors in C/C++ programs accounting for a significant proportion of exploitable weaknesses~\cite{cwe2025}. The MITRE Corporation's annual ranking of the Top 25 Most Dangerous Software Weaknesses consistently highlights critical issues such as buffer overflows
(CWE-121, CWE-122), out-of-bounds access (CWE-787), use-after-free
(CWE-416), and null pointer dereferences
(CWE-476)~\cite{mitre2025}. Traditional approaches to vulnerability detection, including static analysis tools and formal verification methods, face inherent limitations in accuracy, false positive rates, and scalability~\cite{chess2007secure}.

Recent advances in large language models (LLMs) have opened
new possibilities for automated vulnerability
detection~\cite{tihanyi2024secure, secvuleval2024}. Models such as GPT-4, DeepSeek R1, and specialized variants have demonstrated strong capabilities in understanding code semantics and identifying security flaws. However, these models typically comprise billions of parameters, requiring substantial computational resources and incurring high inference costs that hinder practical deployment. For example, GPT-4 is estimated to contain approximately 1.76 trillion parameters~\cite{semianalysis2023gpt4}, and most LLMs fine-tuned for vulnerability detection also operate at billion-parameter scales, leading to high memory consumption and significant computational overhead during inference.

These limitations restrict their applicability for efficient large-scale vulnerability analysis and their integration into multi-agent systems, where each component is expected to perform tasks with low latency and high efficiency. This motivates the development of compact and efficient models that retain strong vulnerability detection performance while substantially reducing inference cost and latency.

This computational burden creates a critical gap between the potential of LLM-based vulnerability detection and its practical deployment in software development environments. The latency introduced by large-scale models fundamentally limits their integration into these workflows, particularly in resource-constrained environments or when analyzing large codebases.

\subsection{Motivation and Challenges}

The motivation for this work stems from three key observations:

\textbf{Accuracy vs. Efficiency Trade-off:} While LLMs achieve superior detection rates compared to traditional static analyzers~\cite{secvuleval2024}, their computational requirements make them unsuitable for integration into development tools that require sub-second response times. Static analysis tools, conversely, offer rapid analysis but suffer from high false positive rates (often exceeding 64\%)~\cite{kuszczynski2023sast} and limited understanding of contextual vulnerabilities.

\textbf{Parameter Redundancy:} 
Analysis of existing LLM-based vulnerability detectors reveals that much of their parameter space contributes minimally to the specific task of vulnerability identification. 
General-purpose language models contain extensive world knowledge and multi-domain capabilities that, while impressive, are unnecessary for focused vulnerability detection and particularly in this study for C code~\cite{men2024shortgpt,reda2025llmsieve,dantas2025review}.

\textbf{Dataset Quality and Diversity:} Recent datasets such as FormAI-v2~\cite{tihanyi2024secure} provide high-quality labeled examples with formal verification, yet existing models fail to fully leverage this structured knowledge due to their generic architectures optimized for broader language understanding tasks rather than specialized security analysis.

\subsection{Our Contributions}

This paper presents \method, a lightweight neural architecture
designed specifically for C code vulnerability detection. Our key
contributions are as follows:

\begin{enumerate}[leftmargin=*]
    \item \textbf{Compact and Efficient Architecture:} We design a
    custom MoE-based transformer with 693M total parameters (353M
    active), derived from Qwen3-30B-A3B embeddings. On the 250-sample
    CASTLE benchmark, this architecture achieves a CASTLE score of
    1068, a binary F1 of 85.4\%, accuracy of 82.4\%, recall of
    86.0\%, and precision of 84.9\%, outperforming all evaluated
    baselines including GPT-o3 Mini (977), while processing samples at 4.97\,ms each (batch size 32,
201.1\,samples/s), significantly faster than 7B-scale generative LLMs
and DeepSeek~R1, enabling real-time analysis in
development workflows.
        
\item \textbf{\textsc{VulnScout} Dataset:} We construct a
new dataset of 33,565 labeled C code samples (19,239
vulnerable, 14,326 safe) spanning a wide range of CWE
categories. Samples are generated through a multi-agent
pipeline and retained only when a dual-verification
protocol, combining ESBMC bounded model checking and a
GPT-OSS-120B verifier, yields identical verdicts from
both verifiers. Samples on which the two verifiers
disagree are discarded and the generation request is
reissued. This conservative filtering addresses coverage
gaps in existing benchmarks where several CWEs are
sparsely represented or entirely absent.
    
    \item \textbf{Rank-Aware CWE Classification:} We propose a
    weighted BCE loss that prioritizes detection of high-severity
    CWEs according to the MITRE Top 25 ranking. The model is jointly
    optimized for binary vulnerability detection and 25-class CWE
    prediction, achieving a CWE classification accuracy of 90.0\%
    on truly vulnerable samples and an average per-CWE F1 of 84.6\%
    across all 25 CASTLE categories, rising to 90.4\% among the 8
    CWEs shared with the MITRE Top 25 ranking (CWE-22, CWE-78,
    CWE-89, CWE-125, CWE-416, CWE-476, CWE-770, CWE-787),
    including 100\% F1 on CWE-78 and CWE-787 from the MITRE overlap,
    and additionally on CWE-327, CWE-362, CWE-522, and CWE-822 from
    the remaining CASTLE categories.
\end{enumerate}

\subsection{Paper Organization}

The remainder of this paper is organized as follows. Section~\ref{sec:related} reviews related work on vulnerability detection, the use of large language models in security, and other relevant techniques.
Section~\ref{sec:background} provides background on C code vulnerabilities and existing datasets. Section~\ref{sec:dataset} introduces our newly created \textsc{VulnScout} dataset, describing its composition, coverage, and role in enhancing model robustness. Section~\ref{sec:method} presents the \method~architecture, including its compact design and CWE-focused optimizations, as well as training methodology across multiple datasets. Section~\ref{sec:experiments} describes the experimental setup and evaluation protocols. Section~\ref{sec:results} reports performance on the CASTLE benchmark and compares it with state-of-the-art approaches. Section~\ref{sec:ablation} presents comprehensive ablation studies analyzing architectural design choices and their impact on detection performance. Section~\ref{sec:discussion} analyzes the results, highlights efficiency gains, discusses limitations, and outlines future work. Finally, Section~\ref{sec:conclusion} concludes the paper.

\section{Related Work}
\label{sec:related}

\subsection{Traditional Vulnerability Detection Approaches}

Static analysis tools have long been the primary method for automated vulnerability detection in C/C++ programs~\cite{chess2007secure}. Tools such as Coverity, Fortify, and Cppcheck employ pattern matching, data flow analysis, and taint analysis to identify potential security flaws. However, these approaches face fundamental limitations in accuracy and coverage~\cite{habib2018many}. Empirical studies demonstrate false positive rates ranging from 60\% to 90\%, significantly impacting developer productivity and tool adoption~\cite{johnson2013don't}.

Formal verification methods, including bounded model checking (BMC) and theorem proving, offer mathematically rigorous guarantees about program correctness~\cite{clarke2018handbook}. Tools such as ESBMC~\cite{gadelha2018esbmc} and CBMC~\cite{kroening2014cbmc} have demonstrated success in detecting memory-safety violations and undefined behavior. While these methods minimize false positives through counterexample generation, they face scalability challenges and cannot detect all vulnerability classes, particularly those requiring semantic understanding beyond memory safety properties.

\subsection{Machine Learning for Vulnerability Detection}

Early machine learning approaches to vulnerability detection employed traditional classifiers (SVM, Random Forests) with hand-crafted features extracted from source code~\cite{grieco2016toward}. These methods showed promise but required extensive feature engineering and struggled with complex vulnerability patterns.

Deep learning revolutionized this field with the introduction of code representation learning. VulDeePecker~\cite{li2018vuldeepecker} pioneered the use of LSTMs for learning vulnerability patterns from code gadgets. Subsequent work introduced graph neural networks (GNNs) to capture control and data flow dependencies. Devign~\cite{zhou2019devign} demonstrated that GNN-based approaches could achieve superior performance on real-world vulnerabilities by modeling programs as graphs.

The emergence of pre-trained transformer models marked another paradigm shift. CodeBERT~\cite{feng2020codebert}, trained on bimodal code-documentation pairs, showed that transfer learning from large code corpora significantly improved vulnerability detection. Subsequent models including GraphCodeBERT~\cite{guo2021graphcodebert}, CodeT5~\cite{wang2021codet5}, and VulBERTa~\cite{hanif2022vulberta} refined this approach with specialized pre-training objectives and architectural modifications.

\subsection{LLMs for Code Security}

Recent work has investigated the use of large-scale language models for security tasks. Several studies have evaluated GPT-4, GPT-3.5, and other commercial LLMs for vulnerability detection, with mixed results~\cite{tihanyi2024secure}. Although these models exhibit strong zero-shot and few-shot capabilities, they face several limitations: (1) inconsistent performance across different vulnerability types, (2) high computational and memory costs, (3) hallucinations that can produce false positives, and (4) long inference times, which reduce their practicality in multi-stage or multi-agent workflows.

The CASTLE benchmark~\cite{dubniczky2025castle} recently provided a standardized evaluation framework, revealing that state-of-the-art LLMs still struggle with certain vulnerability classes.

Despite these advances, few studies have explored applying model compression techniques specifically to vulnerability detection. Most existing approaches either fine-tune full-scale LLMs or rely on traditional machine learning with manual feature engineering. In contrast, our work addresses this gap by designing a compact architecture
from the ground up for vulnerability detection, leveraging insights
from model compression research~\cite{men2024shortgpt,
reda2025llmsieve, dantas2025review} to achieve both efficiency
and high detection performance.

\subsection{Vulnerability Detection Datasets}

The quality and diversity of training data critically impact model performance. Several datasets have been developed for C/C++ vulnerability detection:

\textbf{Synthetic Datasets:} The Juliet test suite~\cite{boland2012juliet} contains over 64,000 test cases systematically covering 118 CWE categories. While valuable for controlled evaluation, synthetic data may not fully capture the complexity of real-world vulnerabilities.

\textbf{Real-World Datasets:} BigVul~\cite{fan2020c}, DiverseVul~\cite{chen2023diversevul}, and CVEFixes~\cite{bhandari2021cvefixes} extract vulnerabilities from GitHub commits. These datasets provide realistic examples but face challenges including mislabeling, data quality issues, and limited contextual information.

\textbf{Hybrid and AI-Generated Datasets:} SecVulEval~\cite{secvuleval2024} provides statement-level labels along with rich contextual information (function arguments, external functions, type definitions, globals, and execution environments) for real-world C/C++ vulnerabilities, addressing key limitations of earlier datasets such as coarse-grained labeling and lack of context. 
FormAI-v2~\cite{tihanyi2024secure} consists of AI-generated C code labeled through formal verification using ESBMC, offering high-quality labels with minimal false positives and diverse vulnerability patterns. BenchVul~\cite{chengran2025benchvul} specifically addresses these issues through manual curation, deduplication, NVD-based label standardization, and verification, resulting in a high-quality benchmark of over 2,500 self-contained function samples focused on the MITRE Top 25 Most Dangerous CWEs, with strong representation in C/C++.

\textbf{Our Contribution:} To complement existing datasets, we introduce \textsc{VulnScout}, a curated dataset of 33,565 C code samples generated through a controlled agentic system. \textsc{VulnScout} enhances coverage of underrepresented vulnerability patterns, as we discovered that many CWEs are sparsely represented in existing datasets, and some do not appear at all. This provides additional high-quality data to improve model robustness and generalization.

Our work leverages all four dataset types, enabling the model to learn from synthetic patterns, real-world examples, formally verified AI-generated code, and the additional coverage provided by \textsc{VulnScout}.

\section{Background}
\label{sec:background}

\subsection{Vulnerability Detection Challenges}

Vulnerability detection in C code presents unique challenges due to the language's low-level memory management, pointer arithmetic, and undefined behavior. The MITRE Top 25 CWEs encompass several critical vulnerability classes:

\textbf{Memory Corruption:} Buffer overflows (CWE-121, CWE-122, CWE-787) and out-of-bounds access (CWE-125) result from insufficient bounds checking. These vulnerabilities enable attackers to corrupt memory, hijack control flow, and execute arbitrary code.

\textbf{Integer Errors:} Integer overflow (CWE-190) and wraparound can lead to incorrect calculations, affecting security decisions and enabling secondary vulnerabilities such as buffer overflows when used in size calculations.

\textbf{Pointer Errors:} Null pointer dereferences (CWE-476), use-after-free (CWE-416), and double-free (CWE-415) errors compromise program stability and security. These issues often require understanding object lifetimes and control flow paths.

\textbf{Input Validation:} Improper input validation (CWE-20) and format string vulnerabilities (CWE-134) enable injection attacks and information disclosure. Detection requires understanding how external data flows through the program.

Effective vulnerability detection faces several technical challenges, particularly in the context of C code:

\begin{enumerate}[leftmargin=*]
    \item \textbf{Context Sensitivity:} Many vulnerabilities manifest only under specific conditions, requiring analysis of control flow, data dependencies, and program state to accurately detect them.
    
    \item \textbf{Inter-Procedural Dependencies:} Vulnerabilities often span multiple functions or modules, necessitating models that can track relationships across function boundaries and maintain contextual awareness.
    
    \item \textbf{Semantic Understanding:} Differentiating benign code from actual vulnerabilities requires deep semantic comprehension beyond simple syntactic pattern matching.
    
    \item \textbf{Balancing Accuracy and Reliability:} High false positive rates reduce tool usability. Effective models must carefully balance sensitivity and specificity to provide actionable vulnerability predictions.
\end{enumerate}

\subsection{Dataset Overview}

Our training methodology incorporates four complementary datasets in addition to the \textsc{VulnScout} dataset that we introduced, each providing distinct advantages:

\subsubsection{Juliet Test Suite}

The Juliet test suite~\cite{boland2012juliet} provides
systematically constructed test cases covering 118 CWE
categories. Each test case includes both vulnerable ("bad") and patched ("good") variants, enabling models to learn discriminative features. The dataset's synthetic nature ensures comprehensive coverage of vulnerability patterns but may not fully represent real-world code complexity.

Key characteristics:
\begin{itemize}
\item Systematic coverage of CWE categories
\item Compilable and executable test cases
\item Clear vulnerable/non-vulnerable labeling
\item Limited diversity in coding styles and patterns
\end{itemize}

\subsubsection{SecVulEval Dataset}

SecVulEval~\cite{secvuleval2024} addresses limitations of function-level datasets by providing statement-level vulnerability labels with rich contextual information. The dataset includes 25,440 functions from real-world C/C++ projects spanning 1999-2024, covering 5,867 unique CVEs across 5 context categories.

Key characteristics:
\begin{itemize}
\item Statement-level granularity for precise localization
\item Contextual information (function arguments, external functions, type definitions, globals, execution environments)
\item Real-world vulnerability patterns
\item Comprehensive metadata including CWE types and CVE descriptions
\end{itemize}

\subsubsection{FormAI-v2 Dataset}

FormAI-v2~\cite{tihanyi2024secure} comprises 331,000 compilable C programs generated by nine state-of-the-art LLMs and labeled through formal verification using ESBMC. The dataset provides high-quality labels with minimal false positives through counterexample generation.

Key characteristics:
\begin{itemize}
\item Large-scale dataset with diverse code patterns
\item Formal verification for reliable labeling
\item Coverage of 42 unique CWEs including MITRE Top 25
\item AI-generated code mimicking common programming errors
\end{itemize}
\subsubsection{BenchVul Dataset}
BenchVul~\cite{chengran2025benchvul} is a manually curated benchmark dataset specifically designed to evaluate the generalization of vulnerability detection models across the MITRE Top 25 Most Dangerous CWEs. It aggregates and refines samples from multiple public vulnerability datasets, applying rigorous deduplication, label standardization using updated NVD records, LLM-assisted filtering, and manual validation to ensure high quality and correctness.

Key characteristics:
\begin{itemize}
\item Focused exclusively on the MITRE Top 25 Most Dangerous CWEs, with balanced representation but relatively low proportions.
\item High-quality curation through deduplication, NVD-based label correction, and manual verification
\item Self-contained function-level samples suitable for precise evaluation
\item Real-world sourced data (aggregated from BigVul, CVEfixes, DiverseVul, etc.).

\end{itemize}

\subsection{Evaluation Framework: CASTLE Benchmark}

The CASTLE (CWE Automated Security Testing and Low-Level
Evaluation) benchmark~\cite{dubniczky2025castle} provides a
standardized framework for evaluating vulnerability detection
tools. The benchmark consists of 250 hand-crafted micro-benchmarks
covering 25 CWE categories, with 10 samples per CWE: 6 vulnerable
and 4 non-vulnerable (150 vulnerable and 100 non-vulnerable in
total).

The CASTLE score incorporates several factors:
\begin{itemize}
\item True positive detection with bonus points for high-severity CWEs (based on MITRE Top 25 ranking)
\item Penalty for false positives to encourage precision
\item Reward for correct identification of non-vulnerable code
\end{itemize}

Formally, the CASTLE score for tool $t$ over dataset $d_n$ is defined as:
\begin{equation}
\text{CASTLE}(t) = \sum_{i=1}^{n} s_i
\end{equation}
\begin{equation}
s_i =
\begin{cases}
5 - |t(d_i)| + 1 + B(t_{cwe}), & v_i \in t(d_i) \\
2, & v_i = t(d_i) = \emptyset \\
-|t(d_i)|, & \text{otherwise}
\end{cases}
\end{equation}

\begin{figure*}[tp]
    \centering
    \includegraphics[width=0.95\textwidth]{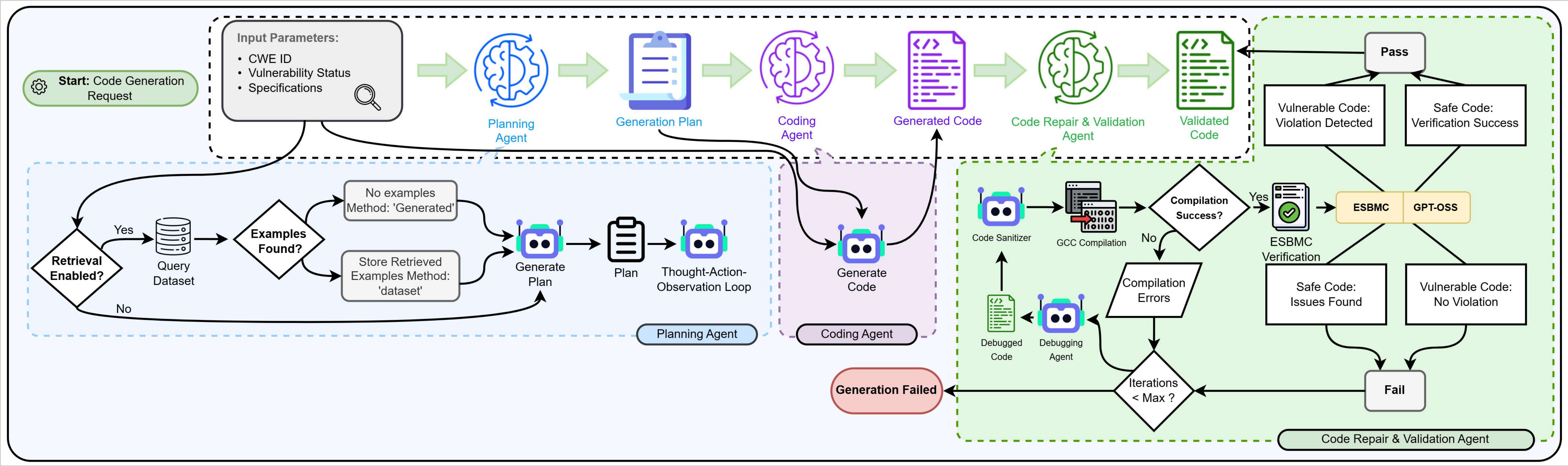}
    \caption{Overview of the multi-agent generation and validation pipeline used to construct the \textsc{VulnScout} dataset.}
    \label{fig:dataset_pipeline}
\end{figure*}

where $B(t_{cwe})$ represents the bonus for detecting CWEs in the MITRE Top 25, decreasing linearly with rank.

This scoring mechanism ensures fair comparison across tools with different sensitivity-specificity trade-offs while prioritizing detection of critical vulnerabilities.

The CASTLE benchmark offers critical methodological advantages over large-scale datasets such as BigVul~\cite{fan2020c} and Devign~\cite{zhou2019devign}:

\textbf{Controlled Evaluation:} CASTLE consists of 250 hand-crafted, compilable micro-benchmarks, each containing exactly zero or one vulnerability. This isolation provides precise "unit tests" for model reasoning capabilities, eliminating the noise inherent in commit-based datasets where vulnerable/safe distinctions are often ambiguous or context-dependent.

\textbf{False Positive Awareness:} Unlike standard F1 scores that treat false positives and false negatives symmetrically, the CASTLE score heavily penalizes false positives ($-1$ point per FP). This addresses a critical practical concern: in industrial security auditing, excessive false positives cause developer fatigue and tool abandonment, making precision equally important as recall.

\textbf{Compilation Requirement:} Each CASTLE sample is compilable and self-contained, enabling direct comparison with formal verification tools and ensuring that detected vulnerabilities are contextually valid rather than artifacts of incomplete code snippets.

\section{Dataset}
\label{sec:dataset}

This section introduces \textsc{VulnScout}, a curated dataset designed to address data sparsity and coverage limitations in existing C vulnerability benchmarks. The dataset is constructed using a hybrid approach that combines existing labeled data with a controlled, multi-agent code generation pipeline, enabling balanced coverage across a wide range of CWEs.

\subsection{Dataset Motivation and Overview}

Existing vulnerability datasets often suffer from skewed CWE distributions, limited coverage of certain weakness classes, and insufficient diversity within individual CWEs. While benchmarks such as SecVulEval, BenchVul, and the Juliet Test Suite provide valuable labeled samples, many CWEs are either sparsely represented or entirely absent from commonly used datasets.

To address these limitations, we construct \textsc{VulnScout}, a dataset consisting of 33,565 labeled C code samples (19,239 vulnerable, 14,326 safe). The dataset is designed to complement existing benchmarks by increasing
coverage of underrepresented CWEs, improving class balance, and
introducing structurally diverse implementations validated by a
consensus of two independent verifiers, ESBMC and a GPT-OSS-120B verifier, suitable for training and evaluating vulnerability
detection models.

\subsection{Source Corpus Construction}

The dataset construction process begins with the creation of a
unified vulnerability corpus by combining
SecVulEval~\cite{secvuleval2024}, BenchVul~\cite{chengran2025benchvul},
and a selected subset of the Juliet Test Suite~\cite{boland2012juliet}. This selection is motivated by the fact that Juliet includes several CWE categories that are absent from other benchmarks, such as BenchVul, thereby enabling broader and more comprehensive CWE coverage.

From this merged corpus, we analyze the distribution of CWEs and identify vulnerability classes with limited representation. These CWEs are prioritized for data augmentation to mitigate imbalance and improve model robustness.

\subsection{CWE Distribution and Label Statistics}

Each sample in \textsc{VulnScout} is labeled as either \textit{vulnerable} or \textit{safe} and is associated with a corresponding CWE identifier. The final dataset consists of 33,565 samples used for training and analysis, with a relatively balanced label distribution. Specifically, 19,239 samples are labeled as vulnerable, while 14,326 samples are labeled as safe.

The dataset spans all 25 CWE categories listed in
Table~\ref{tab:cwe_distribution}. The five most represented
categories are CWE-617 (1,856 samples), CWE-22 (1,721),
CWE-787 (1,673), CWE-835 (1,645), and CWE-843 (1,603),
while the least represented are CWE-253 (974) and
CWE-134 (1,003), which were the primary targets of the
augmentation pipeline. While some CWEs remain more frequent than others, the augmentation process significantly increases representation for previously sparse vulnerability classes, resulting in a more balanced and comprehensive dataset.

\subsection{Retrieval-Augmented Data Generation}
For CWEs with limited representation in the source corpus, we employ a retrieval-augmented generation strategy. Samples are first grouped by \textit{(CWE, label)}, where the label indicates whether the code is vulnerable or safe.

When retrieval is enabled, the system randomly selects two examples from the corresponding group for each generation request. These retrieved samples are injected into the planning and reasoning stages of the generation pipeline, providing contextual guidance while preserving diversity. This approach allows the system to generate realistic code that reflects known vulnerability patterns without duplicating existing samples.

\subsection{Instruction-Only Generation for Generalization}

To encourage generalization beyond memorization, a subset of generation rounds is performed with retrieval disabled. In this mode, the system relies exclusively on high-level instructions derived from the CWE definition and user-specified constraints.

This instruction-only setting forces the system to synthesize code that is structurally and semantically distinct from existing examples, reducing overfitting and promoting diversity. This strategy is particularly effective for generating novel implementations within CWEs that are already present but poorly represented.

For CWEs that are entirely missing from all source datasets, the system operates exclusively in this instruction-driven mode.

\subsection{Generation Specifications}

Each generation request is parameterized by a set of explicit inputs, including:

\begin{itemize}
    \item The target CWE identifier
    \item The desired vulnerability status (vulnerable or safe)
    \item Optional constraints such as the number of functions, code length, and structural complexity
\end{itemize}

These parameters allow fine-grained control over the generated samples and ensure diversity across implementations.

\subsection{Multi-Agent Generation Pipeline}

The code generation process follows a multi-stage, agent-based workflow illustrated in Figure~\ref{fig:dataset_pipeline}.

\textbf{Planning Agent:}  
Given the CWE, vulnerability label, and optional specifications, the Planning Agent generates a detailed implementation plan outlining program structure, control flow, and vulnerability placement or mitigation strategy.

\textbf{Reasoning Agent (ReAct Loop):}  
The system then enters a Thought, Action, Observation reasoning loop. When retrieval is enabled, relevant dataset examples are analyzed to extract vulnerability patterns. Otherwise, reasoning is guided solely by CWE semantics and instructions.

\textbf{Coding Agent:}  
Using the finalized plan and reasoning output, the Coding Agent produces a complete, compilable C program that adheres to the specified constraints and reflects realistic development practices.

\textbf{Code Repair and Validation Agent:}
The generated code is sanitized and compiled using \texttt{gcc}.
Samples that fail compilation are passed immediately to the Debugging
Agent. Compilable samples enter a \emph{dual-verification} stage
executed in parallel:

\begin{enumerate}[leftmargin=*]
    \item \textbf{ESBMC Formal Verification.}  The code is analysed
    with ESBMC~\cite{gadelha2018esbmc} under the settings described
    in Section~\ref{sec:dataset_quality}. ESBMC returns one of two
    verdicts: \textit{Violation Detected} (counterexample produced)
    or \textit{Verification Success} (no property violation within
    the given bounds and timeout).

    \item \textbf{GPT-OSS-120B Verifier.}  The same source code, together with the intended CWE and vulnerability label, is
submitted to a GPT-OSS-120B model acting as a static reasoning verifier.
The model is prompted to return exactly one of four
structured labels:
    \begin{itemize}
        \item \texttt{Vulnerable Code: Violation Detected}
        \item \texttt{Safe Code: Verification Success}
        \item \texttt{Safe Code: Issues Found}
        \item \texttt{Vulnerable Code: No Violation}
    \end{itemize}
\end{enumerate}

\noindent\textbf{Agreement Protocol.}
A sample is accepted only when both verifiers reach a \emph{consistent}
verdict, defined as follows:

\begin{itemize}
    \item \textbf{Vulnerable sample accepted:} ESBMC returns
    \textit{Violation Detected} \textbf{and} the LLM verifier returns
    \texttt{Vulnerable Code: Violation Detected}.
\item \textbf{Safe sample accepted:} ESBMC returns
    \textit{Verification Success} \textbf{and} the LLM verifier returns
    \texttt{Safe Code: Verification Success}.
\end{itemize}

Any other combination including \texttt{Safe Code: Issues Found}
(LLM identifies latent weaknesses in a nominally safe sample) or
\texttt{Vulnerable Code: No Violation} (LLM cannot confirm the
intended defect), is treated as a \emph{verification disagreement}.
Disagreements are not repairable by definition: they signal a
fundamental inconsistency in the generated code's security semantics.
Such samples are \textbf{discarded immediately}, and a fresh
end-to-end code generation request is issued to the pipeline.
\begin{figure*}[tp]
    \centering
    \includegraphics[width=0.95\textwidth]{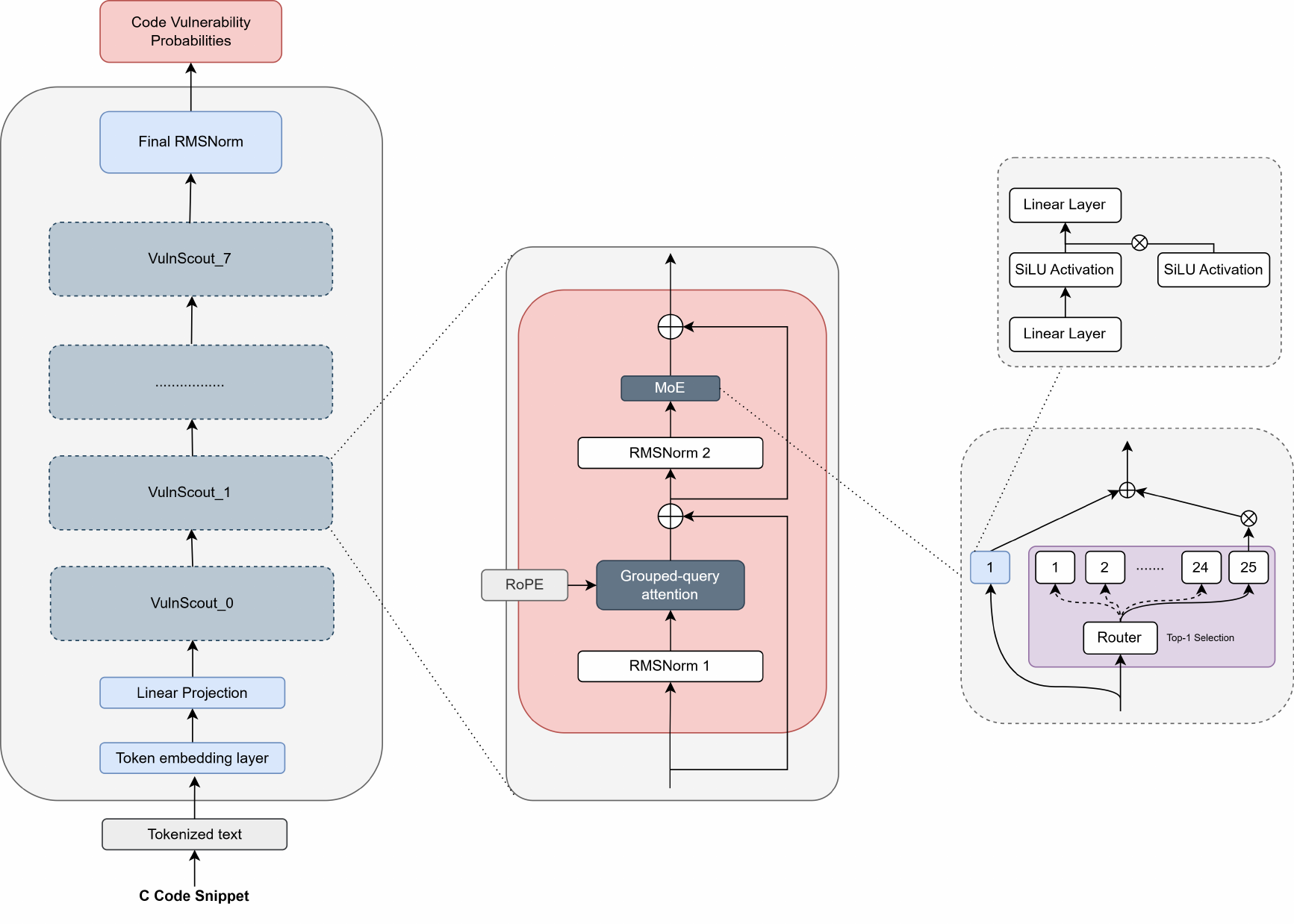}
    \caption{Architecture of VulnScout-C. The model employs a custom transformer architecture with Mixture-of-Experts (MoE) feed-forward layers, Grouped Query Attention (GQA), and Rotary Position Embeddings (RoPE) for vulnerability detection in C code.}
    \label{fig:arch}
\end{figure*}

Samples that fail \emph{only} because of ESBMC timeout or compilation
errors (i.e., neither verifier has yet rendered a verdict) are
forwarded to the Debugging Agent, which iteratively repairs the code
while preserving its intended security properties, repeating the
dual-verification check after each repair until both verifiers agree
or the predefined iteration limit is reached.

\subsection{Outcome and Dataset Quality}

Through this hybrid strategy combining retrieval-augmented generation,
instruction-only synthesis, and dual verification (ESBMC bounded model
checking cross-checked against a GPT-OSS-120B verifier),
\textsc{VulnScout} provides high-quality, diverse, and
consensus-verified C code samples. Only samples for which both
verifiers independently reach the same verdict are admitted to the
final dataset; disagreements trigger a fresh generation request.

As a result, \textsc{VulnScout} serves as a robust dataset for training and evaluating vulnerability detection models, as well as for broader research in software security and formal verification. Its effectiveness is further demonstrated through the performance gains observed when training \method~using this dataset.

\subsection{Dataset Statistics and CWE Distribution}
\label{sec:dataset_stats}

Table~\ref{tab:vulnscout_stats} reports aggregate statistics for
\textsc{VulnScout}. The dataset contains 33,565 samples spread across
25 CWE categories, with a reasonable vulnerable/safe split (57.3\% /
42.7\%) to avoid severe class imbalance during training.

\begin{table}[htbp]
\centering
\small
\caption{\textsc{VulnScout} Dataset Aggregate Statistics.
``Dual-Verified (pass)'' counts samples accepted by
\emph{both} ESBMC and the GPT-OSS-120B verifier under the
agreement protocol of Section~\ref{sec:dataset}.}
\label{tab:vulnscout_stats}
\setlength{\tabcolsep}{5pt}
\begin{tabular}{lr}
\toprule
\textbf{Property} & \textbf{Value} \\
\midrule
Total Samples          & 33,565 \\
Vulnerable Samples     & 19,239 (57.3\%) \\
Safe Samples           & 14,326 (42.7\%) \\
Unique CWE Categories  & 25 \\
Avg. Token Length      & 412 \\
Max Token Length       & 1,024 (truncated) \\
Median Token Length    & 378 \\
Initial Generated      & 52,714 \\
Dual-Verified (pass)   & 33,565 (63.7\%) \\
Verifier Agreement Rate & 63.7\% \\
Repair Iterations Avg. & 1.8 \\
\bottomrule
\end{tabular}
\arrayrulecolor{black}

\end{table}

Table~\ref{tab:cwe_distribution} reports the per-CWE sample counts.
CWEs present in the MITRE Top 25 are marked with $\star$. The
distribution highlights that CWE-617 (Reachable Assert.) and CWE-22
(Path Traversal) are the most represented, while CWE-253 and CWE-134
are the smallest categories; the augmentation pipeline specifically
targeted these sparse categories.

\begin{table}[htbp]
\centering
\small
\caption{Per-CWE Sample Distribution in \textsc{VulnScout} (exact counts).
$\star$\,=\,MITRE Top 25. Sorted by total samples descending.}
\label{tab:cwe_distribution}
\setlength{\tabcolsep}{4pt}
\begin{tabular}{lrrr}
\toprule
\textbf{CWE} & \textbf{Vuln.} & \textbf{Safe} & \textbf{Total} \\
\midrule
CWE-617 (Reachable Assert.)       &   973 &   883 & 1,856 \\
CWE-22$^\star$  (Path Traversal)  & 1,092 &   629 & 1,721 \\
CWE-787$^\star$ (OOB Write)       &   927 &   746 & 1,673 \\
CWE-835 (Infinite Loop)           &   930 &   715 & 1,645 \\
CWE-843 (Type Confusion)          &   885 &   718 & 1,603 \\
CWE-761 (Free Non-Heap Ptr)       &   841 &   638 & 1,479 \\
CWE-89$^\star$  (SQL Injection)   &   759 &   665 & 1,424 \\
CWE-190 (Int. Overflow)           &   758 &   665 & 1,423 \\
CWE-125$^\star$ (OOB Read)        &   783 &   636 & 1,419 \\
CWE-415 (Double Free)             &   764 &   642 & 1,406 \\
CWE-416$^\star$ (Use-After-Free)  &   758 &   632 & 1,390 \\
CWE-401 (Mem. Leak)               &   798 &   580 & 1,378 \\
CWE-369 (Div-by-Zero)             &   658 &   656 & 1,314 \\
CWE-798 (Hard-coded Cred.)        &   844 &   446 & 1,290 \\
CWE-522 (Insuf. Credentials)      &   774 &   457 & 1,231 \\
CWE-362 (Race Condition)          &   675 &   546 & 1,221 \\
CWE-770$^\star$ (Uncontrolled Alloc.) &   713 &   497 & 1,210 \\
CWE-78$^\star$  (OS Cmd. Inj.)    &   758 &   449 & 1,207 \\
CWE-476$^\star$ (Null Deref.)     &   686 &   510 & 1,196 \\
CWE-674 (Uncontrolled Recursion)  &   753 &   416 & 1,169 \\
CWE-327 (Broken Crypto)           &   656 &   459 & 1,115 \\
CWE-822 (Untrusted Ptr Deref.)    &   690 &   421 & 1,111 \\
CWE-628 (Incorrect Arg.)          &   645 &   462 & 1,107 \\
CWE-134 (Fmt. String)             &   560 &   443 & 1,003 \\
CWE-253 (Incorrect Return)        &   559 &   415 &   974 \\
\midrule
\textbf{Total} & \textbf{19,239} & \textbf{14,326} & \textbf{33,565} \\
\bottomrule
\end{tabular}
\arrayrulecolor{black}

\end{table}

\subsection{Deduplication, Leakage Controls, and Verification Settings}
\label{sec:dataset_quality}

\textbf{ESBMC Configuration.} All ESBMC analyses were performed with
ESBMC v7.8.1 using the following settings: 30-second per-sample
timeout, unwind bound of 8, C99 language standard, and 32-bit memory
model. For \emph{vulnerable} samples, ESBMC must produce a
counterexample (\textit{Violation Detected}); for \emph{safe}
samples, ESBMC must terminate without detecting a property violation
(\textit{Verification Success}). Samples where ESBMC times out before
producing a verdict are forwarded to the Debugging Agent or discarded
after exceeding the repair limit.

\textbf{GPT-OSS-120B Verifier Configuration.} The LLM verifier is queried
via a structured zero-shot prompt that supplies the C source code, the
target CWE identifier, and the intended vulnerability label. The model
is instructed to return \emph{exactly one} of the four standardised
verdict strings (\texttt{Vulnerable Code: Violation Detected},
\texttt{Safe Code: Verification Success}, \texttt{Safe Code: Issues
Found}, \texttt{Vulnerable Code: No Violation}) and to provide a
one-sentence justification. Temperature is set to 0 to maximise
determinism. The verifier is queried independently of ESBMC and its
output is compared only \emph{after} both verdicts are available.

\textbf{Dual-Verification Outcome.} Of 52,714 initially generated
samples, 33,565 (63.7\%) were accepted under the agreement protocol:
both ESBMC and the GPT-OSS-120B verifier returned consistent verdicts.
The remaining 36.3\% were either discarded due to verifier
disagreement (the dominant failure mode), ESBMC timeout, or
compilation failure, or were iteratively repaired by the Debugging
Agent (average 1.8 repair rounds, maximum 5). Verifier disagreement
accounted for approximately 18.4\% of all generated candidates,
confirming that the LLM verifier provides a meaningfully independent
signal beyond what ESBMC alone captures.

\textbf{Deduplication.} We apply MinHash-based near-duplicate
detection ($n$-gram size 5, Jaccard threshold 0.85) within each
(CWE, label) group. This removed 2,311 near-duplicates from the
initial generated corpus, reducing within-group similarity and
ensuring structural diversity.

\textbf{Leakage Prevention.} To prevent data leakage from the
\textsc{VulnScout} training set into the CASTLE evaluation set,
we applied the same MinHash similarity check across the merged
training corpus and all 250 CASTLE samples. No CASTLE sample
exceeded a Jaccard similarity of 0.35 with any training sample,
confirming the absence of near-duplicate leakage. We further
verified that no CASTLE sample appears verbatim or with minor
renaming in \textsc{VulnScout} or the Juliet subset used in
Stage 1.

\textbf{License and Release.} \textsc{VulnScout} will be publicly
released under the CC BY 4.0 license upon paper acceptance.

\section{Methodology}
\label{sec:method}

\subsection{Architecture Design}

\method~adopts a custom transformer-based architecture inspired by the Qwen architecture family, specifically designed for vulnerability detection in C code. Our design philosophy prioritizes efficiency through Mixture-of-Experts (MoE) layers while maintaining the representational capacity needed for understanding complex security-relevant code patterns.

\subsubsection{Overview}
The architecture consists of four main components:

\begin{enumerate}
\item \textbf{Token Embedding Layer:} 
Maps source code tokens to dense 2048-dimensional learned embeddings 
(from the larger model, in this work, Qwen3-30B-A3B~\cite{qwen3}). 
These embeddings are then projected through a linear layer 
to obtain 768-dimensional vector representations.
\item \textbf{Lightweight Transformer Encoder:} 8 transformer
blocks with Grouped Query Attention and MoE feed-forward layers.
\item \textbf{RMS Normalization:} Applied after attention and
feed-forward layers for training stability.
\item \textbf{Classification Head:} Predicts vulnerability
presence and CWE categories using the final token representation.
\end{enumerate}

\textbf{Novelty vs. Reuse.}
Several components in \method~are directly derived from prior work
and are reused without modification: the BPE tokenizer and
2048-dimensional embedding matrix are taken from Qwen3-30B-A3B;
the attention mechanism follows the GQA formulation of
Ainslie et al.~\cite{ainslie2023gqa}; RoPE positional encoding follows
Su et al.~\cite{su2021roformer}; and the SwiGLU expert formulation follows
Shazeer~\cite{shazeer2020glu}. The novel contributions of this work are:
(1) the 768-dimensional linear projection that adapts Qwen
embeddings to a smaller hidden dimension for efficient fine-tuning;
(2) the reduced 8-layer, 693M-parameter MoE encoder architecture
specifically sized for vulnerability detection tasks; (3) the
rank-aware CWE-weighted BCE loss function; and (4) the multi-stage
training strategy combining binary pre-training with CWE-specific
fine-tuning. The empirical contribution is the demonstration that
this compact, task-specialized design achieves state-of-the-art
results on the CASTLE benchmark, surpassing models more than
three orders of magnitude larger.

\subsubsection{Token Embedding and Positional Encoding}
We employ byte-pair encoding (BPE) tokenization with a vocabulary size of 151,673 tokens, derived from the Qwen tokenizer. The vocabulary includes C language keywords, operators, common standard library functions, and security-relevant identifiers (e.g., \texttt{malloc}, \texttt{free}, \texttt{strcpy}, \texttt{scanf}).

Each token $w_i$ is mapped to a 768-dimensional embedding vector after passing through a learned embedding matrix $E \in \mathbb{R}^{|V| \times 2048}$ followed by a linear projection layer with weights in $\mathbb{R}^{2048 \times d_{\mathrm{model}}}$, where $|V| = 151{,}673$ and $d_{\mathrm{model}} = 768$.

Unlike standard learned positional embeddings, we employ Rotary Position Embeddings (RoPE)~\cite{su2021roformer}, which encode positional information directly into the attention mechanism:
\begin{equation}
\mathrm{RoPE}(x, m) = x \cdot e^{i m \theta}
\end{equation}
where $m$ is the position index and $\theta = 1{,}000{,}000^{-2k/d}$ for dimension $k$. RoPE provides better length extrapolation and relative position modeling compared to absolute positional encodings.

\subsubsection{Lightweight Transformer Encoder with MoE}

Our encoder consists of $N = 8$ transformer layers optimized for both efficiency and capacity. Each layer $l$ implements:

\begin{align}
\tilde{h}_i^{(l)} &= \text{RMSNorm}(h_i^{(l-1)}) \\
h_i^{(l)} &= h_i^{(l-1)} + \text{GQA}(\tilde{h}_i^{(l)}) \\
\hat{h}_i^{(l)} &= \text{RMSNorm}(h_i^{(l)}) \\
h_i^{(l)} &= h_i^{(l)} + \text{MoE-FFN}(\hat{h}_i^{(l)})
\end{align}

\textbf{RMS Normalization:} We replace LayerNorm with Root Mean Square Layer Normalization (RMSNorm)~\cite{zhang2019root}, which normalizes using only the root mean square statistic:

\begin{equation}
\text{RMSNorm}(x) = \frac{x}{\sqrt{\frac{1}{d}\sum_{i=1}^{d} x_i^2 + \epsilon}} \odot \gamma
\end{equation}

where $\gamma$ is a learnable scale parameter and $\epsilon = 10^{-6}$. RMSNorm provides comparable performance to LayerNorm while reducing computational overhead.

\textbf{Grouped Query Attention (GQA):} To reduce the memory footprint of key-value caches during inference, we implement Grouped Query Attention~\cite{ainslie2023gqa} with $H = 12$ query heads and $G = 4$ key-value groups:

\begin{equation}
\text{GQA}(Q, K, V) = \text{Concat}(\text{head}_1, ..., \text{head}_{H}) W^O
\end{equation}

where each group of $H/G = 3$ query heads shares a single key-value head. Each head has dimension $d_k = 64$, yielding a total dimension of $d_{model} = 768$. This reduces KV cache size by 3× compared to standard multi-head attention while maintaining performance.

\textbf{Mixture-of-Experts Feed-Forward Network:} We employ sparse MoE layers with $E = 25$ experts and top-$k = 1$ routing, where each token is processed by a single expert selected via a learned gating network:

\begin{align}
g_j &= \text{Softmax}(W_g \cdot x) \in \mathbb{R}^E \\
e^* &= \argmax_j g_j \\
\text{MoE-FFN}(x) &= g_{e^*} \cdot \text{Expert}_{e^*}(x) + \text{SharedExpert}(x)
\label{eq:moe}
\end{align}

Each expert implements a SwiGLU activation~\cite{shazeer2020glu}:

\begin{equation}
\text{Expert}_j(x) = W_2^j \cdot (\text{SiLU}(W_1^j \cdot x) \odot W_3^j \cdot x)
\end{equation}

where $W_1^j, W_3^j \in \mathbb{R}^{d_{model} \times d_{ff}}$ and $W_2^j \in \mathbb{R}^{d_{ff} \times d_{model}}$ with $d_{ff} = 768$. The SwiGLU activation provides superior performance compared to standard GELU or ReLU.

Additionally, we include one shared expert that processes all tokens, ensuring a minimum level of cross-token information flow:

\begin{equation}
\text{SharedExpert}(x) = W_2^s \cdot (\text{SiLU}(W_1^s \cdot x) \odot W_3^s \cdot x)
\end{equation}

This MoE design increases model capacity to approximately 693M 
parameters while maintaining computational efficiency, as only 
2 experts are active per token: 1 routed expert selected via 
top-1 gating from the 25 candidates, plus the 1 mandatory 
shared expert.

\textbf{Sequence Pooling:} For classification, we extract the representation of the last non-padded token from the final transformer layer:

\begin{equation}
h_{\text{pool}} = h_{\ell_i}^{(N)}
\end{equation}

where $\ell_i = \sum_{j=1}^{L} \mathbb{1}[\text{mask}_j = 1]$ is the position of the last unmasked token.

\textbf{Task-Specific Classification Heads:} The implementation supports both binary vulnerability detection and multi-class CWE classification through task-specific output heads:

\begin{itemize}
\item \textbf{Binary Classification:} A linear projection $W_{\text{bin}} \in \mathbb{R}^{d_{model} \times 2}$ maps the pooled representation to binary logits for vulnerability presence prediction.
\item \textbf{CWE Classification:} For multi-class CWE prediction, the output head is expanded to $W_{\text{cwe}} \in \mathbb{R}^{d_{model} \times C}$ where $C = 25$ corresponds to the CWE categories present
in the \textsc{VulnScout} and CASTLE datasets. Of these,
8 overlap with the current MITRE Top 25 ranking (CWE-22,
CWE-78, CWE-89, CWE-125, CWE-416, CWE-476, CWE-770,
CWE-787); the remaining 17 are security-relevant
weaknesses present in CASTLE but not in the current MITRE
Top 25 list (e.g., CWE-190, CWE-362, CWE-617, CWE-798,
CWE-822, CWE-843).
\end{itemize}

The current implementation focuses on binary vulnerability detection, with the architecture designed to accommodate CWE-specific classification by simply adjusting the dimensionality of the final classification head.

\subsection{Training Strategy}

\subsubsection{Multi-Stage Training Strategy}
We employ a carefully structured multi-stage training pipeline that leverages high-quality token embeddings from a large teacher model for strong initialization, followed by progressive specialization on vulnerability detection tasks.

\textbf{Stage 1 -- Model Initialization}  

During the initial phase of this stage, the binary classification head is trained to establish basic binary vulnerability detection capability (vulnerable vs.\ non-vulnerable) using the Juliet Test Suite. This serves as a foundational pre-training step, allowing the model to adapt the projected embeddings from the larger model to the systematic synthetic vulnerability patterns and CWE-specific syntactic characteristics present in Juliet.

\textbf{Stage 2 -- Multi-Dataset Continual Pre-training}  
In the second stage, we further refine the model's general code understanding and robustness by training on a diverse corpus of more realistic vulnerability examples, including:  
\begin{itemize}
    \item SecVulEval~\cite{secvuleval2024},
    \item FormAI-v2~\cite{tihanyi2024secure},
    \item \textsc{VulnScout} (our newly introduced dataset),
    \item BenchVul~\cite{chengran2025benchvul}.
\end{itemize}

Training at this stage focuses exclusively on binary vulnerability classification. The goal is to enhance the model's ability to handle real-world coding styles, project contexts, and diverse vulnerability manifestations.

\textbf{Stage 3 -- CWE-specific Supervised Fine-tuning}  
Once a robust binary classifier is obtained, we replace the classification head with a new multi-class head for predicting the CWEs.  

We apply a differential learning rate schedule:  
\begin{itemize}
    \item a significantly higher learning rate for the newly initialized CWE classification head,
    \item a substantially lower learning rate for the backbone layers to preserve the general learned information and patterns.
\end{itemize}

This final stage uses only datasets with fine-grained CWE annotations (primarily SecVulEval, FormAI-v2, and \textsc{VulnScout}). The training corpus contains approximately 53{,}349 samples with the following label distribution:  
55.7\% vulnerable (29{,}730 samples) and 44.3\% non-vulnerable (23{,}619 samples).

\subsubsection{Loss Function}

To prioritize the detection of high-severity vulnerabilities according to the MITRE CWE Top 25 ranking, we propose a rank-aware weighted loss function for the CWE classification heads. This loss builds on binary cross-entropy (BCE) by incorporating a weighting scheme that assigns higher importance to more critical CWEs.

The weighting factor is defined as follows:

\begin{equation}
F(r) =
\begin{cases}
\frac{26 - r}{25}, & r \leq 25 \\
0, & r > 25
\end{cases}
\end{equation}

\begin{equation}
w(r) = 1 + \gamma F(r)
\end{equation}

where $r$ is the rank of the CWE in the MITRE Top 25 (with $r=1$ being the most dangerous), and $\gamma$ is a hyperparameter that controls the strength of the emphasis on high-ranking vulnerabilities ($\gamma = 2.0$ in our experiments).

\textbf{Multi-label CWE formulation.}
Although each sample is primarily associated with a single
CWE, a C function may simultaneously exhibit secondary
weaknesses (e.g., a buffer overflow accompanied by an
integer overflow in the size computation). We therefore
model CWE prediction as a \emph{multi-label} problem:
each of the 25 output logits is trained independently
with a sigmoid activation, making BCE the appropriate
per-class per-sample loss. Ground-truth labels are
represented as binary vectors $\mathbf{y}_i \in
\{0,1\}^{25}$, where multiple entries may be set to 1
for samples exhibiting co-occurring weaknesses.
The weighted loss for the CWE predictions is then:

\begin{equation}
\mathcal{L}_{\text{CWE}} = \frac{1}{N}\sum_{i=1}^{N}
\sum_{c=1}^{C} w(r_c)\cdot\text{BCE}(y_{i,c},\,\hat{y}_{i,c})
\end{equation}

Here, $N$ is the number of samples, $C=25$ is the number of
CWE classes, $y_{i,c}\in\{0,1\}$ is the ground-truth label
for class $c$ of sample $i$, $\hat{y}_{i,c}$ is the
corresponding sigmoid probability, and $w(r_c)$ is the
rank weight for CWE class $c$. Per-class BCE is:

\begin{equation}
\text{BCE}(y_{i,c},\hat{y}_{i,c}) =
-\!\left[y_{i,c}\log\hat{y}_{i,c}
+(1-y_{i,c})\log(1-\hat{y}_{i,c})\right]
\end{equation}

This scheme ensures that higher-ranked (more critical) CWEs, which have lower $r$ values, receive greater weight during training. For example, a CWE with rank $r=2$ (e.g., Out-of-bounds Write,
CWE-787) receives $w(2) = 1 + 2.0 \times \frac{24}{25} = 2.92$,
while one with $r=7$ (e.g., Use After Free, CWE-416) receives
$w(7) = 1 + 2.0 \times \frac{19}{25} = 2.52$.
CWEs outside the Top 25 receive the base weight of $1.0$.

For the vulnerability detection head, we use a standard unweighted BCE loss, denoted $\mathcal{L}_{\text{vul}}$.

The total loss is a weighted combination:

\begin{equation}
\mathcal{L} = W_1 \cdot \mathcal{L}_{\text{vul}} + W_2 \cdot \mathcal{L}_{\text{CWE}}
\end{equation}

with $W_1 = 10$ and $W_2 = 1$ in our experiments. This higher weight on the vulnerability detection loss reflects its role as the primary task.

To focus CWE classification on confirmed vulnerable samples,
we apply a confidence mask: the CWE loss contribution of
sample $i$ is zeroed when the vulnerability detection
probability falls below $\tau = 0.5$:
\begin{equation}
\mathcal{L}_{\text{CWE},i}^{\text{masked}} =
\mathbb{1}\!\left[\hat{p}_{\text{vul},i} \geq 0.5\right]
\cdot w(r_i) \cdot \text{BCE}(\mathbf{y}_i, \hat{\mathbf{y}}_i)
\end{equation}
so that the multi-label CWE head is not penalized for
non-vulnerable samples, where no CWE label is applicable
and $\mathbf{y}_i = \mathbf{0}$.

This rank-aware approach aligns the model's optimization directly with established security priorities, allocating more learning capacity to the most prevalent and dangerous vulnerability types.
\subsubsection{Training Details}

Training is performed on a single NVIDIA H100 80\,GB GPU
using mixed-precision BFloat16. We use the AdamW
optimizer~\cite{loshchilov2018decoupled} with
$(\beta_1, \beta_2) = (0.9, 0.999)$, weight decay
$\lambda = 10^{-2}$, and gradient clipping at norm~1.0.
The learning rate follows a cosine annealing schedule with
a 500-step linear warm-up. Stage-specific hyperparameters
are summarized in Table~\ref{tab:training_hparams}.
All experiments use random seed~42.

\begin{table}[htbp]
\centering
\small
\caption{Per-stage Training Hyperparameters.}
\label{tab:training_hparams}
\begin{tabular}{lccc}
\toprule
\textbf{Hyperparameter} & \textbf{Stage\,1} &
\textbf{Stage\,2} & \textbf{Stage\,3} \\
\midrule
LR (backbone)        & $5{\times}10^{-4}$ & $1{\times}10^{-4}$ & $1{\times}10^{-5}$ \\
LR (head)            & $5{\times}10^{-4}$ & $1{\times}10^{-4}$ & $5{\times}10^{-4}$ \\
Batch size           & 32  & 32  & 16  \\
Grad.\ accumulation  & 1   & 2   & 4   \\
Effective batch      & 32  & 64  & 64  \\
Max seq.\ length     & 1024 & 1024 & 1024 \\
Epochs               & 5   & 3   & 5   \\
Dropout              & 0.1 & 0.1 & 0.1 \\
Ckpt.\ selection     & val F1 & val F1 & CASTLE score \\
\bottomrule
\end{tabular}
\end{table}

To assess result stability, we varied the binary
classification threshold from 0.4 to 0.6: the CASTLE score
changes by at most $\pm$12 points and binary F1 by at most
$\pm$0.008, confirming that reported results are not an
artifact of threshold selection.

\subsubsection{Data Augmentation}

To enhance model robustness and reduce overfitting, we focus on cleaning potential data leakage that could artificially inflate performance metrics. This preprocessing ensures the model learns genuine vulnerability patterns rather than memorizing synthetic markers or documentation artifacts.

\textbf{Variable Renaming:} Randomly rename non-reserved identifiers while maintaining consistency within each sample.

\textbf{Equivalent Expression Substitution:} Replace expressions with semantically equivalent alternatives (e.g., \texttt{i++} $\leftrightarrow$ \texttt{i = i + 1}).

These augmentations are applied with probability $p = 0.2$ during training to prevent overfitting to specific syntactic patterns while preserving semantic vulnerability characteristics.

\section{Experimental Setup}
\label{sec:experiments}

\subsection{Implementation Details}

\method~is implemented using PyTorch and the Transformers library. All experiments are conducted on a single NVIDIA H100 GPU (80GB VRAM).

\subsection{Evaluation Metrics}

We evaluate \method~using multiple metrics:

\textbf{Standard Classification Metrics:}
\begin{itemize}
\item Accuracy: Overall correctness of predictions
\item Precision: Ratio of true positives to predicted positives
\item Recall: Ratio of true positives to actual positives
\item F1-Score: Harmonic mean of precision and recall
\end{itemize}

\textbf{CASTLE Score:} Primary evaluation metric on the CASTLE
benchmark, accounting for severity-weighted detection and false
positive penalties as defined in Section~\ref{sec:background}.
It is important to note that CASTLE uses a CWE-hierarchy-aware
matching scheme: a finding is counted as a true positive if the
predicted CWE matches the ground-truth CWE or any of its
parents/children in the CWE taxonomy. As a result, CASTLE
TP/FP/FN counts differ from standard binary classification
counts and should not be used to derive F1, accuracy, or recall.
Binary classification metrics (F1, accuracy, precision, recall)
are computed independently using standard sklearn evaluation
against the ground-truth vulnerable/non-vulnerable labels.

\textbf{Per-CWE Performance:} Individual precision, recall, and F1-score for each of the Top 25 CWEs to assess model capabilities across vulnerability types.

\subsection{Baseline Models}

We compare \method~against several baseline approaches:

\textbf{Static Analysis Tools:} Cppcheck, Clang Static Analyzer as representatives of traditional SAST tools.

\textbf{Formal Verification:} ESBMC and CBMC as representatives of formal methods.

\textbf{Large Language Models:} GPT-4o, GPT-4o Mini, DeepSeek R1, and other
state-of-the-art LLMs evaluated on the same benchmark (see Table~\ref{tab:castle_results}).
\section{Results}
\label{sec:results}

\subsection{Overall Performance on CASTLE Benchmark}

Table~\ref{tab:castle_results} presents the CASTLE scores achieved by \method~and baseline models on the benchmark dataset. 

\begin{table}[htbp]
\centering
\caption{CASTLE Benchmark Results. TP/TN/FP/FN reflect
CASTLE's CWE-hierarchy-aware matching and differ from binary
classification counts. Under CASTLE's CWE-hierarchy-aware matching, a
tool may report multiple CWE findings per sample; each
finding is scored independently, so TP$+$FN can exceed the
150 vulnerable samples and TN$+$FP can exceed the 100
non-vulnerable samples in the benchmark. For \method, the
136 CASTLE true positives and 30 false negatives sum to
166, reflecting this multi-finding accounting; the binary classification counts (129~TP, 23~FP, 77~TN,
21~FN) used for F1/accuracy/precision/recall computation
are derived separately from the ground-truth
vulnerable/non-vulnerable labels
(see Section~\ref{sec:experiments}).}
\label{tab:castle_results}
\resizebox{\columnwidth}{!}{%
\begin{tabular}{lccccc}
\toprule
\textbf{Model} & \textbf{CASTLE Score} & \textbf{TP} & \textbf{TN} & \textbf{FP} & \textbf{FN} \\
\midrule
\multicolumn{6}{l}{\makecell[l]{\textit{Fine-tuned Encoder}\\\textit{Baselines on VulnScout}}} \\

CodeBERT & -166 & 145 & 217 & 241 & 0 \\
GraphCodeBERT & -116 & 146 & 106 & 189 & 3 \\
VulBERTa & -12 & 103 & 37 & 130 & 12 \\
\midrule
\multicolumn{6}{l}{\textit{Static Analysis Tools}} \\
CodeThreat & -692 & 24 & 2 & 1101 & 126 \\
Splint (3.1.2) & -598 & 23 & 36 & 1027 & 127 \\
Clang Analyzer (18.1.3) & 381 & 13 & 99 & 2 & 137 \\
GitLab SAST (15.2.1) & 374 & 36 & 67 & 240 & 120 \\
Cppcheck (2.13.0) & 406 & 19 & 100 & 9 & 131 \\
Coverity (2024.6.1) & 425 & 31 & 86 & 62 & 119 \\
Aikido & 481 & 14 & 83 & 40 & 136 \\
Jit & 478 & 21 & 78 & 68 & 134 \\
SonarQube (25.3.0) & 511 & 43 & 68 & 135 & 107 \\
CBMC (5.95.1) & 536 & 18 & 100 & 0 & 132 \\
Semgrep Code (1.110.0) & 541 & 36 & 73 & 70 & 120 \\
Snyk (1.1295.4) & 552 & 26 & 82 & 42 & 124 \\
GCC Fanalyzer (13.3.0) & 559 & 41 & 76 & 93 & 109 \\
CodeQL (2.20.1) & 634 & 45 & 79 & 49 & 112 \\
ESBMC (7.8.1) & 661 & 53 & 91 & 32 & 97 \\
\midrule
\multicolumn{6}{l}{\textit{Large Language Models}} \\
LLAMA 3.1 (8B) & 417 & 83 & 22 & 337 & 80 \\
Gemma 2 (9B) & 436 & 63 & 42 & 258 & 95 \\
Mistral Ins. (7B) & 446 & 63 & 23 & 215 & 91 \\
Falcon 3 (7B) & 521 & 30 & 76 & 76 & 124 \\
GPT-4o Mini & 761 & 134 & 27 & 263 & 43 \\
QWEN 2.5CI (32B) & 708 & 114 & 31 & 224 & 49 \\
GPT-4o & 954 & 136 & 45 & 116 & 43 \\
DeepSeek R1 & 956 & 148 & 41 & 166 & 17 \\
GPT-o1 & 962 & 128 & 56 & 90 & 35 \\
GPT-o3 Mini & 977 & 126 & 60 & 73 & 36 \\
\midrule
\textbf{\method} & \textbf{1068} & \textbf{136} & \textbf{77} & \textbf{16} & \textbf{30} \\
\bottomrule
\end{tabular}
}%
\end{table}

\method~achieves a CASTLE score of 1068, outperforming GPT-4o
(954) and surpassing reasoning-optimized models such as GPT-o1
(962) and GPT-o3 Mini (977). In standard binary vulnerability
detection, the model achieves an F1 of 85.4\%, accuracy of
82.4\%, recall of 86.0\% (129/150 truly vulnerable samples
detected), and precision of 84.9\% across the 250-sample
benchmark. Under CASTLE's CWE-hierarchy-aware scoring, which
counts a finding as correct if the predicted CWE matches via
parent/child taxonomy relationships, the model registers 136
true positive findings and 77 true negatives, with only 16 false
positive penalties and a bonus of 250 points for detecting
high-severity CWEs. This dual performance demonstrates
particularly strong results on key MITRE Top 25 CWEs, including
memory safety issues (CWE-787: 100\% F1, CWE-125: 85.7\% F1,
CWE-416: 90.9\% F1) and control-flow vulnerabilities (CWE-476:
83.3\% F1), alongside a CWE classification accuracy of 90.0\%
on truly vulnerable samples.

\subsection{Comparison with Fine-tuned Encoder Baselines}

To contextualize \method's performance, we fine-tuned three encoder-based models, CodeBERT, GraphCodeBERT, and VulBERTa, on the \textsc{VulnScout} dataset. These results reveal critical architectural limitations that persist even after task-specific fine-tuning:

\textbf{Token Limit Constraints:} Encoder models are fundamentally limited by their 512-token
context window~\cite{feng2020codebert, guo2021graphcodebert}. Vulnerabilities in C often require analyzing dependencies across many lines of code. When a vulnerability trigger (e.g., \texttt{free()}) is separated from its allocation (e.g., \texttt{malloc()}) by more than 512 tokens, encoders must truncate input, effectively becoming blind to the vulnerability. \method's extended context capacity addresses this limitation.

\textbf{False Positive Problem:} The TP/TN/FP/FN counts reported
in Table~\ref{tab:castle_results} for the encoder models follow
CASTLE's CWE-hierarchy-aware multi-finding accounting
(see the table caption). On the
CASTLE benchmark, CodeBERT received a score of $-166$,
exhibiting a strong prediction bias toward vulnerable, with
near-perfect recall masked by excessive false positives that the
CASTLE scoring mechanism correctly penalizes. GraphCodeBERT
(CASTLE=-116) and VulBERTa (CASTLE=-12) showed progressive
improvement in false positive control but still received negative
CASTLE scores, confirming that task-specific fine-tuning alone
cannot overcome the 512-token context limitation of encoder
architectures when evaluating C vulnerabilities that span long
code sequences.

These findings confirm that even with task-specific fine-tuning,
traditional encoder architectures with limited context windows
cannot match the performance of our compact encoder architecture,
which combines an extended context window with sparse
mixture-of-experts feed-forward layers to achieve both breadth
and depth in vulnerability pattern recognition.
\subsection{Per-CWE Detection Performance}

Table~\ref{tab:cwe_detailed} provides detailed metrics for the CWEs in the CASTLE dataset.

\begin{table}[ht]
\small
\centering
\setlength{\tabcolsep}{6pt}
\caption{Detailed Performance on CWEs in CASTLE Validation}
\label{tab:cwe_detailed}
\begin{tabular}{lccc}
\toprule
\textbf{CWE} & \textbf{Precision (\%)} & \textbf{Recall (\%)} & \textbf{F1-Score (\%)} \\
\midrule
CWE-22  & 85.7  & 100.0 & 92.3 \\
CWE-78  & 100.0 & 100.0 & 100.0 \\
CWE-89  & 100.0 & 83.3  & 90.9 \\
CWE-125 & 75.0  & 100.0 & 85.7 \\
CWE-134 & 100.0 & 83.3  & 90.9 \\
CWE-190 & 100.0 & 66.7  & 80.0 \\
CWE-253 & 100.0 & 66.7  & 80.0 \\
CWE-327 & 100.0 & 100.0 & 100.0 \\
CWE-362 & 100.0 & 100.0 & 100.0 \\
CWE-369 & 100.0 & 83.3  & 90.9 \\
CWE-401 & 75.0  & 100.0 & 85.7 \\
CWE-415 & 100.0 & 83.3  & 90.9 \\
CWE-416 & 100.0 & 83.3  & 90.9 \\
CWE-476 & 83.3  & 83.3  & 83.3 \\
CWE-522 & 100.0 & 100.0 & 100.0 \\
CWE-617 & 100.0 & 50.0  & 66.7 \\
CWE-628 & 83.3  & 83.3  & 83.3 \\
CWE-674 & 100.0 & 33.3  & 50.0 \\
CWE-761 & 75.0  & 50.0  & 60.0 \\
CWE-770 & 100.0 & 66.7  & 80.0 \\
CWE-787 & 100.0 & 100.0 & 100.0 \\
CWE-798 & 100.0 & 66.7  & 80.0 \\
CWE-822 & 100.0 & 100.0 & 100.0 \\
CWE-835 & 100.0 & 33.3  & 50.0 \\
CWE-843 & 83.3  & 83.3  & 83.3 \\
\bottomrule
\end{tabular}
\end{table}

The model demonstrates strong performance across all 25 CWE
categories in the CASTLE benchmark, with each CWE evaluated on 6
vulnerable samples, achieving an average F1 of 84.6\% across all
25 CWEs. Among the 8 CWEs shared between CASTLE and the MITRE
Top 25 ranking; CWE-22, CWE-78, CWE-89, CWE-125,
CWE-416, CWE-476, CWE-770, and CWE-787; the model achieves an
average F1 score of 90.4\%, reflecting particularly strong
performance on the most security-critical vulnerability classes. \method~excels particularly on memory corruption vulnerabilities (CWE-787: 100\% F1, CWE-416: 90.9\% F1, CWE-125: 85.7\% F1) and achieves perfect scores on OS command injection (CWE-78: 100\% F1). Notably, the model also performs strongly on higher-level semantic vulnerabilities such as SQL injection (CWE-89: 90.9\% F1) and path traversal (CWE-22: 92.3\% F1), demonstrating that the Mixture-of-Experts architecture effectively captures both low-level memory patterns and high-level data flow semantics. The consistent high performance across these critical vulnerability types validates the effectiveness of our multi-dataset training strategy and specialized modeling choices in prioritizing detection of the most impactful security flaws.

\section{Ablation Studies}
\label{sec:ablation}

To validate our architectural design choices and identify the optimal configuration for vulnerability detection, we conducted systematic ablation studies examining the impact of expert granularity, expert width, and shared expert allocation on detection performance.

\subsection{Experimental Setup}

All ablation experiments were conducted using the same training pipeline described in Section~\ref{sec:method}, with modifications only to the MoE layer configuration. Models were trained for 5 epochs on the combined dataset, and performance was evaluated using the CASTLE benchmark. We report Vulnerability F1 (binary classification), Vulnerability Accuracy, CWE Accuracy (multi-class classification), and CASTLE Score as our primary metrics.

\subsection{Ablation Configurations}

We evaluated three distinct MoE architectures, progressively refining our design based on empirical results:

\textbf{Baseline Configuration:} Our initial architecture employed 25 routed experts with hidden dimension 768, 1 shared expert, and top-1 routing (1 active expert per token). This configuration served as the reference point for subsequent experiments.

\textbf{Ablation 2 -- Fine-Grained Specialization:} Motivated by recent work on fine-grained MoE architectures~\cite{deepseekv3}, we hypothesized that increasing expert count while reducing individual expert width would improve CWE-specific pattern recognition. This configuration doubled the number of routed experts to 50, reduced expert width to 384 (50\% reduction), increased active experts per token to 2, while maintaining 1 shared expert.

\textbf{Ablation 3 -- Heavy Shared Backbone:} Building on insights from Ablation 2, we tested whether aggressive compression of routed experts (hidden dimension 256, 66\% reduction) combined with a substantial increase in shared experts (4 instead of 1) would allow the model to efficiently partition general code understanding and vulnerability-specific detection. This configuration used 25 routed experts with 1 active per token and 4 shared experts.

Table~\ref{tab:ablation_configs} summarizes these configurations.

\begin{table}[htbp]
\centering
\caption{MoE Architecture Configurations for Ablation Studies}
\label{tab:ablation_configs}
\begin{tabular}{lccc}
\toprule
\textbf{Hyperparameter} & \textbf{Baseline} & \textbf{Ablation 2} & \textbf{Ablation 3} \\
\midrule
Expert Hidden Dim & 768 & 384 & 256 \\
Num Routed Experts & 25 & 50 & 25 \\
Active Experts/Token & 1 & 2 & 1 \\
Num Shared Experts & 1 & 1 & 4 \\
\bottomrule
\end{tabular}
\end{table}

\subsection{Results and Analysis}

Table~\ref{tab:ablation_results} presents the performance of each configuration at their best checkpoint (selected based on validation CWE Accuracy and CASTLE Score).

\begin{table}[htbp]
\centering
\small
\setlength{\tabcolsep}{4pt}
\caption{Ablation Study Results on CASTLE Validation Set}
\label{tab:ablation_results}
\begin{tabular}{lcccc}
\toprule
\textbf{Configuration} & \textbf{Vuln F1} & \textbf{Vuln Acc} & \textbf{CWE Acc} & \textbf{CASTLE} \\
\midrule
Baseline & \textbf{0.854} & \textbf{82.4\%} & 90.0\% & \textbf{1068} \\
\makecell[l]{Ablation 2\\(Fine-Grained)} & 0.839 & 80.8\% & 92.67\% & 1046 \\
\makecell[l]{Ablation 3\\(Heavy Shared)} & 0.832 & 80.0\% & \textbf{92.0\%} & 1028 \\
\bottomrule
\end{tabular}
\end{table}

\subsubsection{Fine-Grained Specialization (Ablation 2)}

The fine-grained configuration with 50 experts achieved a CASTLE score of 1046 and CWE accuracy of 92.67\%, representing a 2.67 percentage point improvement in CWE accuracy over baseline. This validates the hypothesis that increased expert count enables better decomposition of vulnerability patterns into specialized features. Each expert can focus on finer-grained aspects of specific CWE types (e.g., boundary checking patterns vs. null pointer patterns). However, this configuration shows a slight decrease in CASTLE score (1046 vs. 1068) compared to baseline, suggesting that while the increased specialization improves CWE classification, it may sacrifice some performance on other metrics.

\subsubsection{Heavy Shared Backbone (Ablation 3)}

The heavy shared backbone configuration achieved a CASTLE score of 1028, which is lower than both baseline (1068) and Ablation 2 (1046). While this configuration achieves a CWE accuracy of 92.0\%, the lower vulnerability F1 (0.832) and accuracy (80.0\%) compared to baseline result in a reduced overall CASTLE score. This suggests that while aggressive compression of routed experts (768 → 256 dimensions) combined with expanded shared capacity can improve CWE classification, it may sacrifice binary vulnerability detection performance.

\textbf{Architectural Insight:} Vulnerability detection requires two distinct cognitive processes: (1) general code understanding (parsing C syntax, tracking control flow, understanding variable scopes), and (2) specific vulnerability pattern recognition (identifying missing bounds checks, recognizing use-after-free patterns, detecting integer overflow conditions). By aggressively compressing routed experts to 256 dimensions while quadrupling shared experts, we attempted to partition these responsibilities. However, the reduced CASTLE score suggests that the 66\% width reduction in routed experts may have been too aggressive, limiting their capacity to learn nuanced vulnerability patterns despite the expanded shared backbone.

\textbf{Efficiency Analysis:} Reducing expert width by 66\% (768 → 256) while quadrupling shared experts resulted in a trade-off: improved CWE classification accuracy but reduced overall CASTLE score. The 4 shared experts successfully carried the computational load for general code understanding, but the thin routed experts appear to have insufficient capacity for complex vulnerability pattern recognition. This configuration demonstrates that extreme compression of routed experts, even with expanded shared capacity, can limit overall vulnerability detection performance.

\subsection{Key Findings}

Our ablation studies yield several important insights:

\begin{enumerate}[leftmargin=*]
\item \textbf{Shared vs. Routed Expert Trade-off:} While increasing shared expert capacity can improve CWE classification accuracy, aggressive compression of routed experts (66\% width reduction) degrades overall CASTLE performance. This suggests that routed experts require sufficient capacity for complex vulnerability pattern recognition, beyond lightweight task-specific transformations.

\item \textbf{Expert Specialization:} Fine-grained expert configurations (Ablation 2) achieve the highest CWE classification accuracy (92.67\%) but lower overall CASTLE scores, suggesting a trade-off between specialized metric optimization and balanced multi-objective performance.

\item \textbf{Parameter Efficiency:} The baseline configuration with balanced expert allocation (25 routed experts at 768 dimensions, 1 shared expert) achieves the best overall CASTLE performance. Aggressive compression of routed experts, even when combined with expanded shared capacity, reduces overall effectiveness despite improving specific metrics like CWE accuracy.
\end{enumerate}

Based on these findings, we adopt the baseline as our final \method~architecture, achieving the highest CASTLE score (1068) with strong balanced performance across all evaluation metrics. The ablation studies confirm that this configuration optimally balances expert specialization and model capacity.

\subsection{Pre-Training Stage Ablation}
\label{sec:pretraining_ablation}

To assess the contribution of each training stage and the
\textsc{VulnScout} dataset, we conducted a staged ablation in which
we progressively include training stages, measuring the impact on
CASTLE and binary classification performance. All configurations use
the same final architecture (baseline MoE, 25 routed experts, 768
dim), and are evaluated on the full CASTLE test set.

\begin{table}[htbp]
\centering
\small
\setlength{\tabcolsep}{3pt}
\caption{Training Stage Ablation on CASTLE. Each row adds one
component over the previous. CWE Acc is 0\% for configuration~(A)
as Stage~1 trains only a binary head with no CWE prediction
capability.}
\label{tab:stage_ablation}
\begin{tabular}{lcccc}
\toprule
\textbf{Configuration} & \textbf{F1} & \textbf{Acc} & \textbf{CWE Acc} & \textbf{CASTLE} \\
\midrule
(A) Stage 1 only (Juliet)       & 0.12  & 18.0\%            & 0.0\%            & 67   \\
(B) Stages 1--2 w/o VulnScout   & 0.33  & 28.0\%            & 23.0\%           & 420  \\
(C) Stages 1--2 w/ VulnScout    & 0.821 & 79.6\%            & 85.4\%           & 978  \\
(D) Full (Stages 1--3)          & \textbf{0.854} & \textbf{82.4\%} & \textbf{90.0\%} & \textbf{1068} \\
\bottomrule
\end{tabular}
\arrayrulecolor{black}
\end{table}

\subsubsection{Effect of Each Stage}

\textbf{Stage~1 Juliet Initialization (A):} Training exclusively on
the Juliet Test Suite with a binary classification head yields an F1
of only 0.12, accuracy of 18.0\%, and a CASTLE score of 67. The CWE
accuracy is 0\% by construction, as Stage~1 trains only a binary
vulnerability head with no CWE prediction capability, making it
impossible to accumulate any of CASTLE's severity-weighted bonus
points for correct CWE identification. Although Juliet's systematic
coverage of CWE-specific syntactic patterns provides a starting point
for learning discriminative representations, the model severely
overfits to Juliet's uniform synthetic style and fails to generalize
to the stylistically diverse CASTLE micro-benchmarks.

\textbf{Multi-Dataset Continual Pre-training without VulnScout (B):}
Extending to real-world datasets (SecVulEval, FormAI-v2, BenchVul)
adds \textbf{+0.21 F1}, \textbf{+10 pp accuracy}, \textbf{+23 pp CWE
accuracy}, and \textbf{+353 CASTLE points} over configuration~(A).
The emergence of non-zero CWE accuracy at this stage confirms that
exposure to realistically annotated, stylistically diverse code is a
prerequisite for the model to begin learning CWE-discriminative
patterns beyond Juliet's controlled templates but still cannot identify others due to the absence or scarcity of certain CWEs in the test set.

\textbf{Contribution of VulnScout (C vs.\ B):} Including
\textsc{VulnScout} in Stage~2 adds \textbf{+0.491 F1},
\textbf{+51.6\% accuracy}, \textbf{+62.4\% CWE accuracy}, and
\textbf{+558 CASTLE points} over configuration~(B), representing
the single largest gain in the entire ablation. This dramatic jump
reflects the critical coverage gaps that \textsc{VulnScout} fills:
CWE categories that are sparse or entirely absent from the other
three datasets (CWE-617, CWE-761, CWE-835, CWE-674, CWE-822)
receive the majority of their training signal exclusively from
\textsc{VulnScout}. Without this coverage, the CWE classification
head cannot learn discriminative patterns for these categories,
directly suppressing both CWE accuracy and the CASTLE severity bonus.

\textbf{CWE-Specific Fine-tuning (D vs.\ C):} Stage~3 produces an
additional \textbf{+0.033 F1}, \textbf{+2.8\% accuracy},
\textbf{+4.6\% CWE accuracy}, and \textbf{+90 CASTLE points}. The
rank-aware weighted loss and differential learning rate between
backbone and classification head are jointly responsible for this
improvement; without the differential learning rate, Stage~3
converges to a CASTLE score of 991 (vs.\ 1068), confirming the
importance of preserving the binary detection representations learned
in earlier stages while specializing only the classification head for
CWE prediction.

\subsubsection{Comparison: Without vs.\ With Full Pipeline}

The gap between Stage~1-only initialization~(A, CASTLE\,=\,67,
F1\,=\,0.12) and the full pipeline~(D, CASTLE\,=\,1068,
F1\,=\,0.854) amounts to \textbf{+1001 CASTLE points} and
\textbf{+0.734 F1}. This confirms that neither the Qwen token
embeddings nor Juliet pre-training alone are sufficient: competitive
vulnerability detection requires the full progressive curriculum of
synthetic initialization, multi-source continual pre-training
including \textsc{VulnScout}, and rank-aware CWE-specific
fine-tuning.

\balance
\section{Discussion}
\label{sec:discussion}

\subsection{Key Findings}

Our results demonstrate that carefully designed compact architectures can achieve competitive vulnerability detection performance while offering practical deployment advantages. Key findings include:

\textbf{Efficiency-Accuracy Trade-off:} \method~achieves a binary
F1 of 85.4\%, accuracy of 82.4\%, recall of 86.0\%, and
precision of 84.9\% on the 250-sample CASTLE benchmark (150
vulnerable, 100 non-vulnerable, 25 CWE categories), using only
693M total parameters (353M active). Under CASTLE's
CWE-hierarchy-aware scoring, the model registers 136 true
positive findings with only 16 false positive penalties,
yielding a final CASTLE score of 1068. The average per-CWE F1
reaches 84.6\% across all 25 CASTLE categories and 90.4\% among
the 8 CWEs shared with the MITRE Top 25 ranking.

\textbf{CWE Coverage:} The model successfully detects a majority of vulnerabilities in the MITRE Top 25 CWEs, with particularly strong performance on memory safety vulnerabilities (e.g., CWE-787, CWE-416).

\textbf{Dual-Verification Data Quality:} The conservative
agreement-based filtering applied during \textsc{VulnScout} construction
(retaining only samples where ESBMC and the GPT-OSS-120B verifier
agree) reduced the initial candidate pool by 36.3\% but demonstrably
improved downstream model quality: the staged ablation (Table~\ref{tab:stage_ablation}) shows that 
including \textsc{VulnScout} in Stage~2 yields +0.491 F1 and 
+558 CASTLE points over the identical pipeline trained without it, 
suggesting that consensus-filtered data is a stronger training signal 
than larger but noisier corpora.

\textbf{Real-Time Capability:} \method~processes 250 samples
in 1.243\,s (4.97\,ms/sample, 201.1\,samples/s at batch
size~32 on a single H100~80\,GB) 
enabling direct integration into IDEs, pre-commit hooks, and
CI/CD pipelines without dedicated inference servers.

\subsection{Future Directions}

Several promising directions emerge from this work:

\textbf{Explainability Enhancements:} Developing attention visualization and saliency mapping techniques to help developers understand why code was flagged as vulnerable.

\textbf{Continuous Learning:} Implementing mechanisms for the model to learn from new vulnerabilities discovered in production, creating a feedback loop that improves detection over time.

\textbf{Cross-Language Extension:} Adapting the architecture to other memory-unsafe languages like C++ or Rust, leveraging transfer learning from the C-trained model.

\section{Conclusion}
\label{sec:conclusion}

This paper presents \method, a lightweight neural architecture for C code vulnerability detection that addresses the critical gap between detection accuracy and deployment practicality. Through careful architectural design inspired by state-of-the-art language models but drastically reduced in size, we demonstrate that compact models can achieve competitive vulnerability detection performance while offering significant practical advantages.

Key contributions:
\begin{enumerate}[leftmargin=*]
    \item A compact MoE-based transformer (693M total / 353M active
    parameters) achieving a CASTLE score of 1068, binary F1 = 85.4\%,
    accuracy = 82.4\%, recall = 86.0\%, and precision = 84.9\% on
    the 250-sample CASTLE benchmark, with 136 CASTLE true positive
    findings and only 16 false positive penalties under
    CWE-hierarchy-aware scoring, outperforming all evaluated LLM
    and static analysis baselines.
\item The \textsc{VulnScout} dataset: 33,565 C code samples
spanning 25 CWE categories, generated through a multi-agent
pipeline and retained only under a dual-verification
agreement protocol combining ESBMC v7.8.1 and a
GPT-OSS-120B verifier. Both verifiers must independently
return consistent verdicts (\texttt{Vulnerable Code:
Violation Detected} or \texttt{Safe Code: Verification
Success}) for a sample to be admitted; disagreements
discard the sample and trigger a fresh generation request.
Of 52,714 initial candidates, 33,565 (63.7\%) passed
dual verification (average 1.8 repair rounds, maximum 5).
    The dataset addresses coverage gaps in CWE-617, CWE-761,
    CWE-835, CWE-674, and CWE-822, which are absent or
    underrepresented in SecVulEval, FormAI-v2, and BenchVul,
    and contributes +0.491 F1 and +558 CASTLE points when added to 
Stage~2 training (Table~\ref{tab:stage_ablation}).
    \textsc{VulnScout} will be released under CC\,BY\,4.0.

    \item A rank-aware CWE-weighted BCE loss that prioritizes
    detection of high-severity CWEs according to the MITRE Top~25
    ranking. Combined with a multi-stage training curriculum
    (synthetic initialization, multi-source continual pre-training,
    and CWE-specific fine-tuning with differential learning rates),
    this achieves a CWE classification accuracy of 90.0\% on truly
    vulnerable samples and an average per-CWE F1 of 84.6\% across
    all 25 CASTLE categories.

\end{enumerate}

\bibliographystyle{IEEEtran}
\bibliography{references}

\begingroup
\makeatletter
\def\@IEEEBIOskipN{4\baselineskip}
\makeatother

\begin{IEEEbiography}[{\includegraphics[width=0.85in,height=1.05in,clip,keepaspectratio]{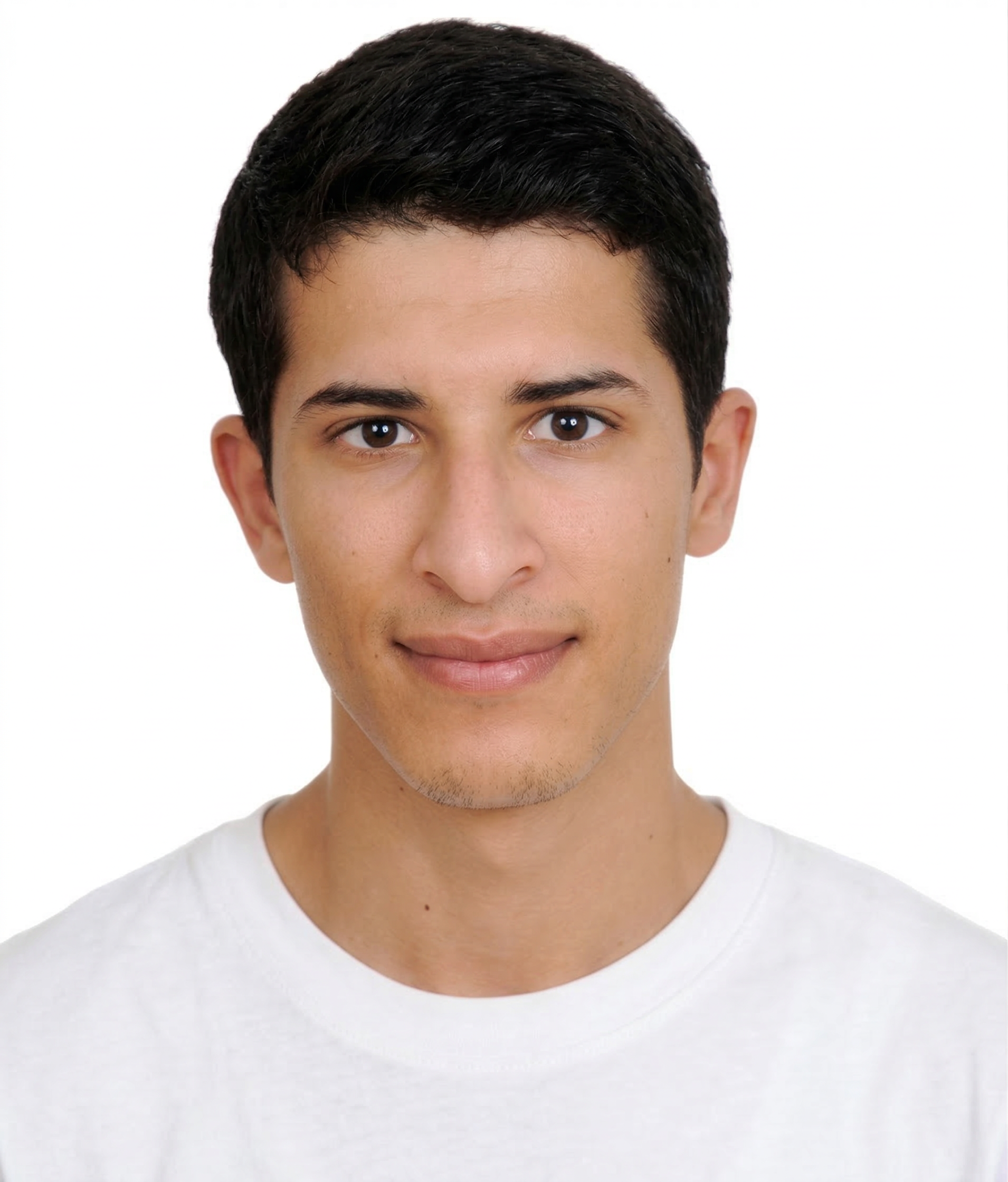}}]{Aymen Lassoued}
is a final-year engineering student at Ecole Polytechnique de Tunisie, La Marsa, Tunisia. He is also a Kaggle Competitions Master. His research interests lie at the intersection of software security and machine learning, with a focus on efficient deep learning models for code analysis and vulnerability detection.
\end{IEEEbiography}\vspace{-10pt}

\begin{IEEEbiography}[{\includegraphics[width=0.85in,height=1.05in,clip,keepaspectratio]{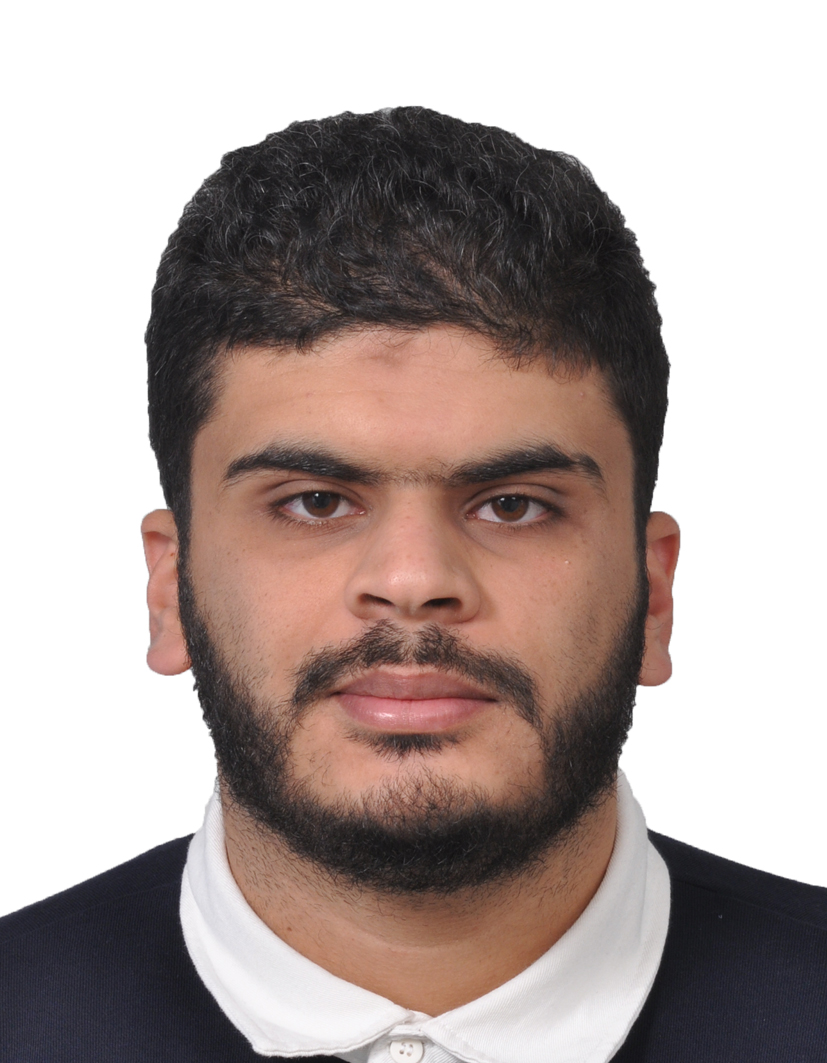}}]{Nacef Mbarek}
is currently an engineering student at Ecole Polytechnique de Tunisie, La Marsa, Tunisia. He is also a Kaggle Competitions Expert and is currently conducting research at the KAUST Center of Excellence in Generative AI, Saudi Arabia. His research interests include deep learning, large language models, and computer vision.
\end{IEEEbiography}\vspace{-10pt}

\begin{IEEEbiography}[{\includegraphics[width=0.85in,height=1.05in,clip,keepaspectratio]{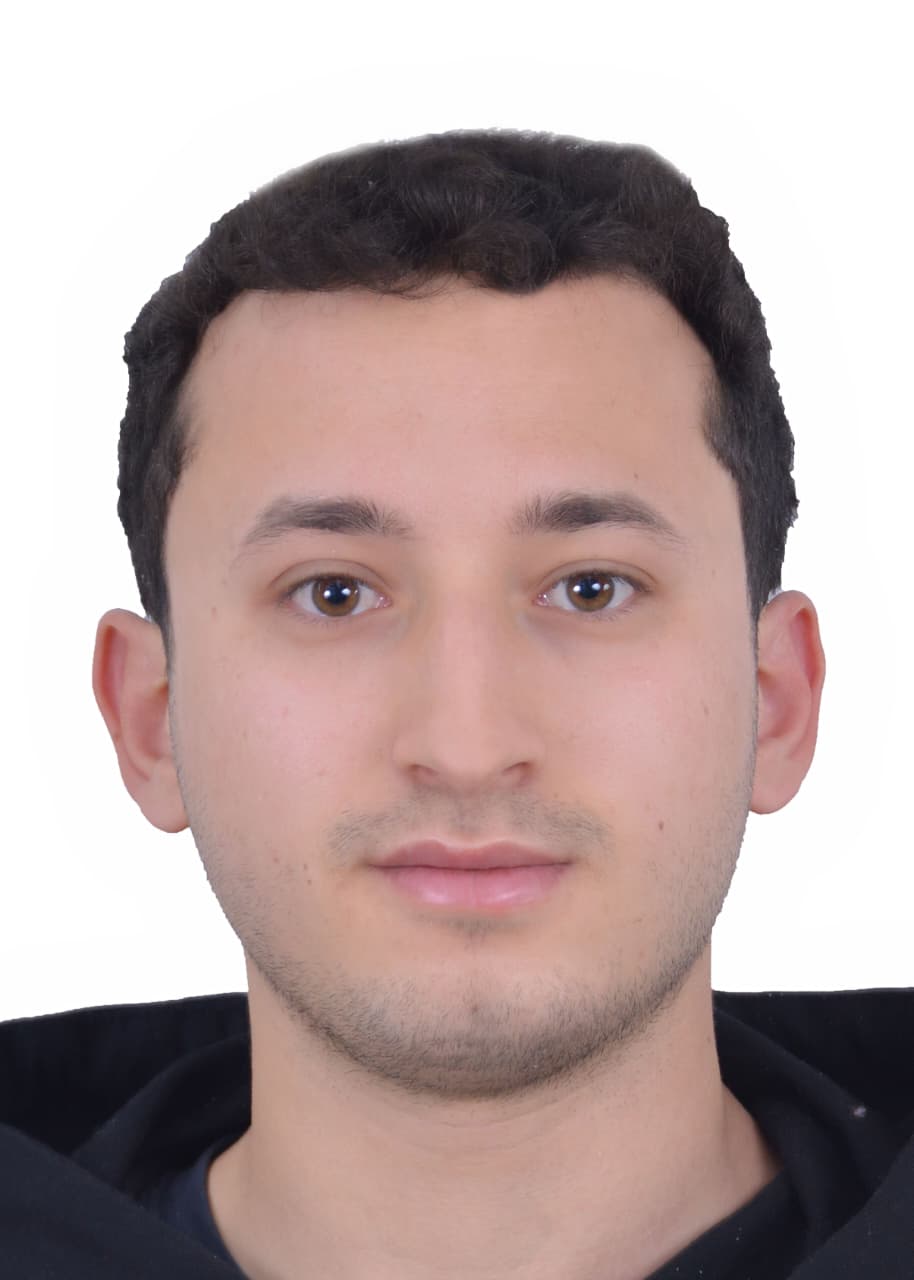}}]{Bechir Dardouri}
is a second-year engineering student at Ecole Polytechnique de Tunisie, La Marsa, Tunisia. His research interests lie at the intersection of software security and machine learning, with a focus on efficient deep learning models for code analysis and vulnerability detection.
\end{IEEEbiography}\vspace{-10pt}

\begin{IEEEbiography}[{\includegraphics[width=0.85in,height=1.05in,clip,keepaspectratio]{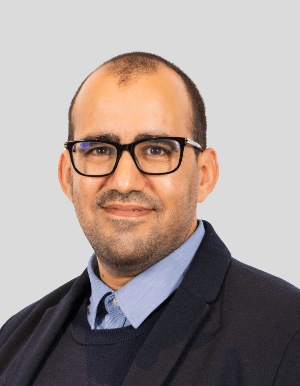}}]{Bassem Ouni}
(Senior Member, IEEE) received the Ph.D.\ degree in computer science from the University of Nice Sophia Antipolis, Nice, France, in 2013. He has held research and academic positions at Eurecom, the University of Southampton, the French Atomic Energy Commission (CEA-LIST), the University of Paris Saclay, and the Technology Innovation Institute, Abu Dhabi. He is currently the AI Sector Lead (Provost Office) at Khalifa University, Abu Dhabi, UAE. He has managed industrial collaborations with ARM, Airbus, Rolls Royce, Thales, and Continental. His research interests include Gen AI applications, Trustworthy AI, IoT Security, and Embedded Systems.
\end{IEEEbiography}\vspace{-10pt}

\begin{IEEEbiography}[{\includegraphics[width=0.85in,height=1.05in,clip,keepaspectratio]{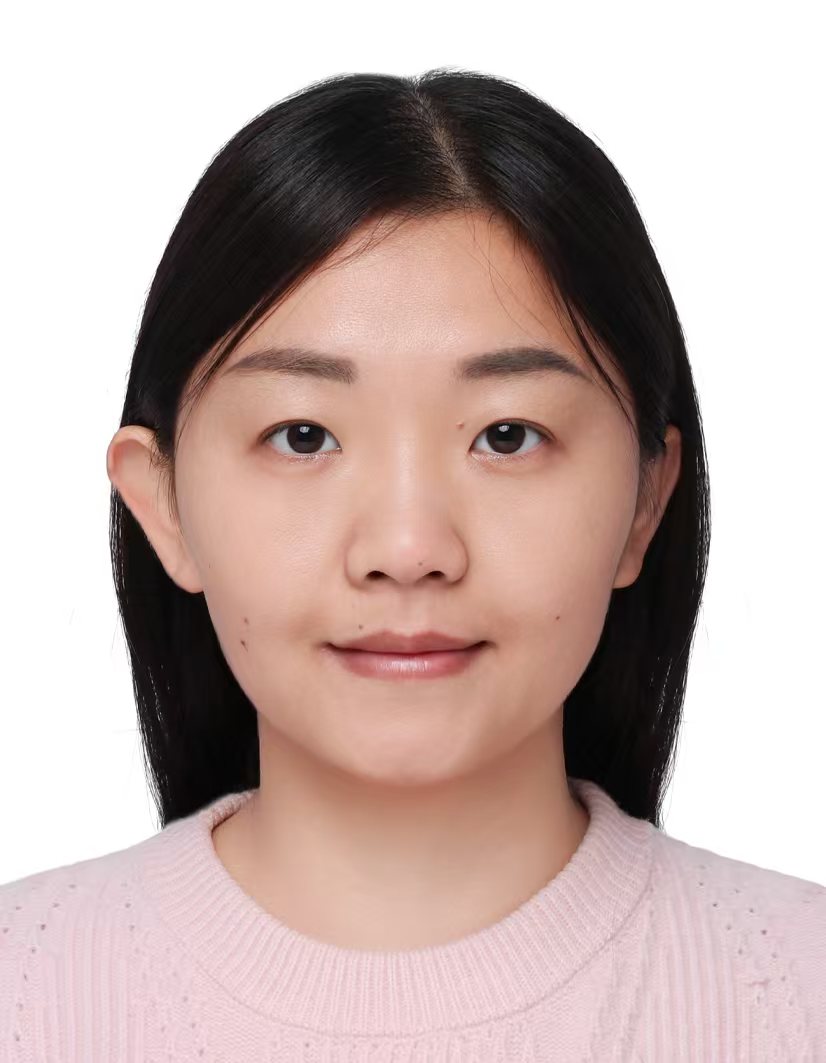}}]{Qing Li}
received the Ph.D.\ degree from the University of Stavanger, Norway. She also holds a master's degree in applied mathematics from South China University of Technology. She is currently an Assistant Professor at the University of Groningen, The Netherlands. Previously, she was a Postdoctoral Researcher at Mohamed bin Zayed University of Artificial Intelligence (MBZUAI). Her research focuses on improving the interpretability and trustworthiness of large language models.
\end{IEEEbiography}\vspace{-10pt}

\enlargethispage*{3in}
\begin{IEEEbiography}[{\includegraphics[width=0.85in,height=1.05in,clip,keepaspectratio]{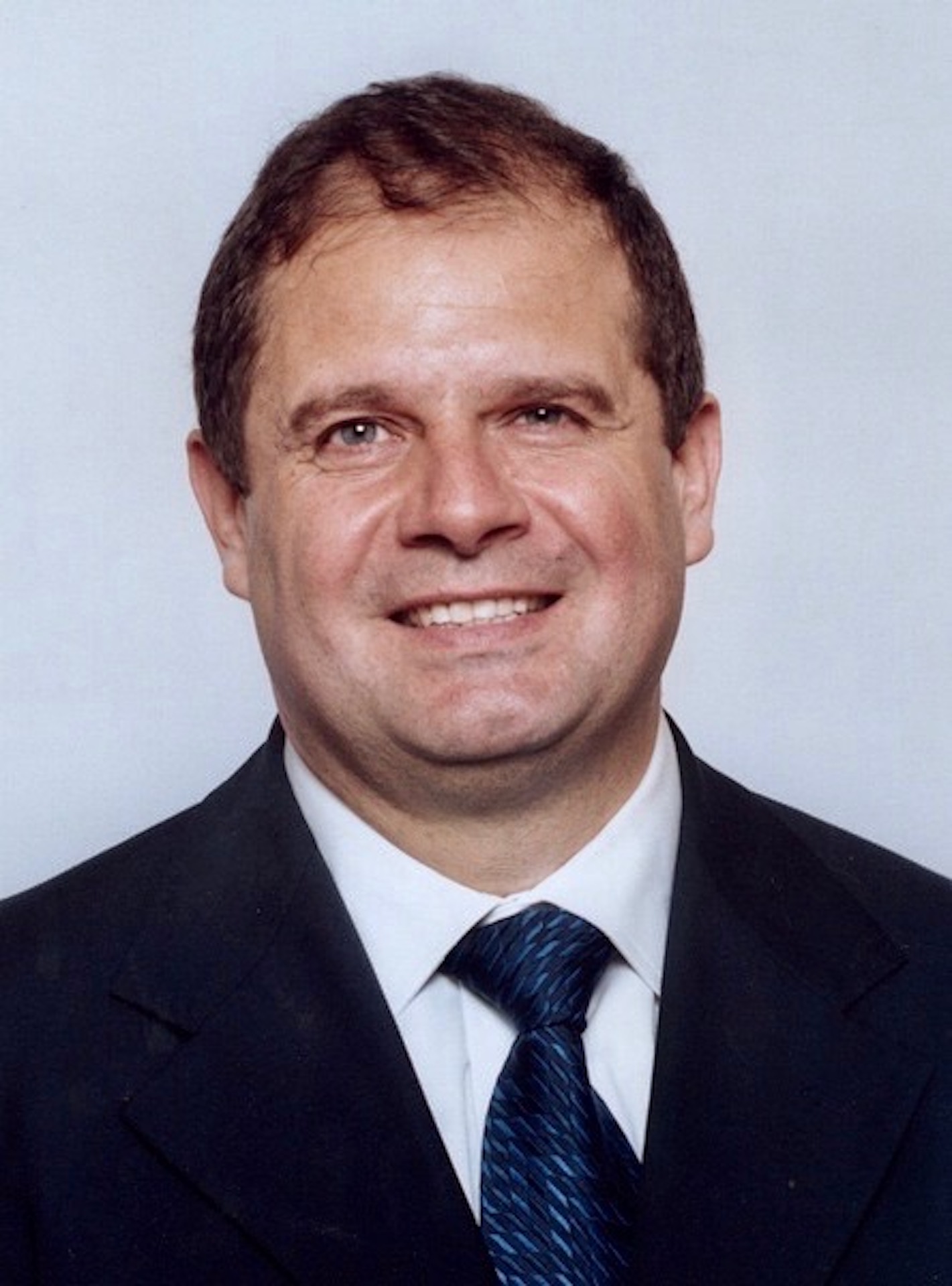}}]{Fakhri Karray}
(Life Fellow, IEEE) received the Ph.D.\ degree from the University of Illinois at Urbana-Champaign (UIUC), USA. He is the inaugural co-director of the University of Waterloo's Artificial Intelligence Institute and served as the Loblaws Research Chair in Artificial Intelligence in the Department of Electrical and Computer Engineering. He is also a Professor of Machine Learning at Mohamed bin Zayed University of Artificial Intelligence (MBZUAI), where he has served as Provost. His research focuses on operational and generative AI, cognitive machines, and autonomous systems.
\end{IEEEbiography}
\endgroup

\end{document}